\documentclass[prd,superscriptaddress,amsfonts,amssymb,amsmath,showpacs,twocolumn]{revtex4-2}
\usepackage{bm}
\usepackage{amsfonts}
\usepackage{latexsym}
\usepackage[latin1]{inputenc}
\usepackage{graphicx}
\usepackage{amsmath,eqnarray}
\usepackage{palatino}
\usepackage{mathpazo}
\usepackage{textcomp}
\usepackage{float}
\usepackage{booktabs}
\usepackage{dcolumn}
\usepackage{hyperref}
\usepackage{mathtools,xcolor}
\hypersetup{colorlinks,citecolor=red}
\usepackage{tikz,caption}

\usepackage{amsmath}
\usepackage{xcolor}
\usepackage{orcidlink}
\usepackage[caption=false]{subfig}
\usepackage{commath}
\definecolor{bluE}{HTML}{8fddf7}
\def\jnl@style{\it}
\def\aaref@jnl#1{{\jnl@style#1}}

\def\aaref@jnl#1{{\jnl@style#1}}

\def\aj{\aaref@jnl{AJ}}                   
\def\apj{\aaref@jnl{ApJ}}                 
\def\apjl{\aaref@jnl{ApJ}}                
\def\apjs{\aaref@jnl{ApJS}}               
\def\apss{\aaref@jnl{Ap\&SS}}             
\def\aap{\aaref@jnl{A\&A}}                
\def\aapr{\aaref@jnl{A\&A~Rev.}}          
\def\aaps{\aaref@jnl{A\&AS}}              
\def\mnras{\aaref@jnl{Mon.~Not.~Roy.~Astron.~Soc.}}             
\def\prd{\aaref@jnl{Phys.~Rev.~D}}        
\def\prc{\aaref@jnl{Phys.~Rev.~C}}  
\def\prl{\aaref@jnl{Phys.~Rev.~Lett.}}    
\def\qjras{\aaref@jnl{QJRAS}}             
\def\skytel{\aaref@jnl{S\&T}}             
\def\ssr{\aaref@jnl{Space~Sci.~Rev.}}     
\def\zap{\aaref@jnl{ZAp}}                 
\def\nat{\aaref@jnl{Nature}}              
\def\aplett{\aaref@jnl{Astrophys.~Lett.}} 
\def\apspr{\aaref@jnl{Astrophys.~Space~Phys.~Res.}} 
\def\physrep{\aaref@jnl{Phys.~Rep.}}      
\def\physscr{\aaref@jnl{Phys.~Scr}}       
\def\commat{\aaref@jnl{Comm.~Math.~Phys.}}              
\def\science{\aaref@jnl{Science}}               
\def\cqg{\aaref@jnl{Classical Quant.~Grav.}}            
\def\jpcs{\aaref@jnl{JPCS}}                                     
\def\ijmpd{\aaref@jnl{Int.~J.~Mod.~Phys.~D}}                    
\def\grg{\aaref@jnl{Gen.~Relat.~Gravit.}}               
\def\rpp{\aaref@jnl{Rep.~Prog.~Phys.}}          
\def\npa{\aaref@jnl{Nucl.~Phys.~A}}        
\def\lrr{\aaref@jnl{Living Rev.~Rel.}}                   
\def\jcap{\aaref@jnl{J.~Cosmology Astropart.~Phys.}}    
\def\rmp{\aaref@jnl{Rev.~Mod.~Phys.}}   
\def\epjc{\aaref@jnl{Eur.~Phys.~J.~C}}

\allowdisplaybreaks[1]

\begin{document}

\color{black}

\title{Wormhole Geometries Supported by Strange Quark Matter and Phantom-like Generalized Chaplygin gas within $f(Q)$ Gravity}

\author{Sneha Pradhan\orcidlink{0000-0002-3223-4085}}
\email{snehapradhan2211@gmail.com}

\author{Zinnat Hassan\orcidlink{0000-0002-3223-4085}}
\email{zinnathassan980@gmail.com}
\affiliation{Department of Mathematics, Birla Institute of Technology and
Science-Pilani,\\ Hyderabad Campus, Hyderabad-500078, India.}

\author{P.K. Sahoo\orcidlink{0000-0003-2130-8832}}
\email{pksahoo@hyderabad.bits-pilani.ac.in}
\affiliation{Department of Mathematics, Birla Institute of Technology and
Science-Pilani,\\ Hyderabad Campus, Hyderabad-500078, India.}

\date{\today}

\begin{abstract}
A crucial aspect of wormhole (WH) physics is the inclusion of exotic matter, which requires violating the null energy condition. Here, we explore the potential for WHs to be sustained by quark matter under conditions of extreme density along with the phantom-like generalized cosmic Chaplygin gas (GCCG) in symmetric teleparallel gravity. Theoretical and experimental studies on baryon structures indicate that strange quark matter, composed of u (up), d (down), and s (strange) quarks, represents the most energy-efficient form of baryonic matter. Drawing from these theoretical insights, we use the Massachusetts Institute of Technology (MIT) bag model equation of state to characterize ordinary quark matter. By formulating specific configurations for the bag parameter, we develop several WH models corresponding to different shape functions for the isotropic and anisotropic cases. Our analysis strongly suggests that an isotropic WH is not theoretically possible. Furthermore, we investigate traversable WH solutions utilizing a phantom-like GCCG, examining their feasibility. This equation of state, capable of violating the null energy condition, can elucidate late-time cosmic acceleration through various beneficial parameters. In this framework, we derive WH solutions for both constant and variable redshift functions. We have employed the volume integral quantifier (VIQ) method for both studies to assess the quantity of exotic matter. Furthermore, we have done the equilibrium analysis through the Tolman-Oppenheimer-Volkoff (TOV) equation, which supports the viability of our constructed WH model. 
\end{abstract}
\maketitle

\section{Introduction}\label{sec:1} 
The principles of general relativity, along with various modified theories, suggest the existence of complex structures within spacetime, one of which includes WHs. These entities act as tunnels, creating pathways that link different or far-off regions of the universe. Within the realm of general relativity, phenomena such as black holes and WHs are particularly captivating subjects of study. While the presence of black holes has been verified through various studies such as Refs. \cite{Abbott1,Abbott2,Abbott3}, the exploration into WHs still continues. The idea of WHs was first introduced by Einstein and Rosen, leading to the concept known as the Einstein-Rosen bridge \cite{Rosen1}. Interest in WHs experienced a revival with Ellis's introduction of a new solution to Einstein's equations, which involved a spherically symmetric setup and a massless scalar field with unusual properties \cite{Rosen2}. Following this, Morris and Thorne \cite{Thorne/1988} illustrated that the Ellis WHs could be traversable, opening up discussions on their potential for enabling swift space travel and possibly even time travel. These models of WHs are noted for lacking singularities or horizons, and they possess tidal forces that are considered effortless for human safety. Furthermore, Morris and Thorne \cite{Thorne/1988} pointed out that these WHs would necessitate the existence of exotic matter due to their violation of the null energy conditions. This exotic matter, which does not adhere to usual energy conditions, has properties that disobey conventional physics, such as the possibility of particles having negative mass. Significant research has been done into the existence and stability of WHs. For instance, Shinkai and Hayward \cite{Khatsymovsky2} used numerical simulations to demonstrate the instability of Ellis WHs. The idea of creating such passageways using ordinary matter has intrigued researchers, especially with recent studies suggesting that in the realm of modified theories of gravity, WHs made from ordinary matter that meet all energy conditions might be possible \cite{Khatsymovsky3}. However, even in these scenarios, the effective geometric matter acting as the source for the modified gravity might still not conform to the traditional null energy condition. Several investigations have also been directed towards WH models that do not rely on exotic matter \cite{Tanaka5,Tanaka6,Tanaka7,Tanaka8}. Apart from that, some beautiful work on WH could be found in the literature \cite{o1,o2,o3}.\\
General Relativity (GR) stands as the prevailing gravitational theory, supported by several experiments and observations. Nonetheless, it fails to explain several phenomena, including the Universe's accelerated expansion, gravitational behavior on the scale of galaxies, and the quest for a quantum gravity theory. To address these challenges, scientists have proposed alternative gravitational models. Among these, the $f(R)$ gravity model is a significant adaptation of GR, where the Einstein-Hilbert action is modified by substituting the Ricci scalar with a function of the scalar curvature \cite{Starobinsky1,Starobinsky2,Starobinsky3}. This adjustment modifies the gravitational field equations, potentially impacting how gravity operates across various scales. The $f(R)$ gravity model has been applied to explain the Universe's accelerated expansion, align with the constraints of early universe inflation, and offer an alternative to dark matter in explaining the rotational behavior of galaxies, as well as aspects of stellar dynamics and galaxy shape \cite{Starobinsky4}.\\
Over recent years, there has been a notable development in the studies related to extension GR \cite{Laurentis}, including torsion-based gravity \cite{Krssak}. However, a significant development occurred in 1999 with the introduction of the Symmetric Teleparallel Gravity concept \cite{Nester,Kalay,Conroy}, which promotes a modified gravity approach where both curvature and torsion are omitted. Instead, this approach attributes gravitational phenomena to the non-metricity tensor, specifically through the non-metricity scalar $Q$. The concept of $f(Q)$ gravity proposed by Jimenez and colleagues \cite{Jimenez} focuses solely on the non-metricity scalar $Q$ to describe the gravitational field. It has gained considerable attention for its potential to explain the Universe's accelerated expansion with statistical accuracy compared to other established modified gravity theories \cite{Lin1}. Extensive research has delved into the cosmological impacts of $f(Q)$ gravity \cite{Lin2,Lin4,Lin5}, showcasing its applications in understanding cosmological evolution \cite{Lin3}, the dynamics of black holes \cite{Shaun1,Shaun2,Shaun3}, and the characteristics of compact stars through gravitational decoupling \cite{Shaun4}. Recent studies have also explored the compact object in the framework of $f(Q)$ gravity \cite{Shaun5,sp1,sp2,sp3}. Also, the solutions for static and spherically symmetric scenarios involving anisotropic fluids within the $f(Q)$ gravity framework are presented \cite{Wang1}. Additionally, the $f(Q)$ theory has been applied to examine the WH geometries, indicating that linear models of $f(Q)$ gravity might minimize the need for exotic matter to form traversable WHs \cite{Hassan}. Recent applications of the theory include its use in exploring the properties of Casimir WHs \cite{Ghosh1}, including those corrected by the Generalized Uncertainty Principle (GUP) \cite{Ghosh2}. Moreover, one can also refer to some interesting literature on astrophysical objects found in non-metricity-based modified theories of gravity (see Refs. \cite{Pradhan1,Pradhan2,Pradhan3,Pradhan4,Pradhan5,Pradhan6,Pradhan7}).\\
\indent In the field of astrophysics, models for stellar compact objects can include strange quark matter (SQM) using the MIT Bag model. This model outlines the relationship as $p = \omega(\rho - 4B)$, where $B$ represents the Bag constant. The coefficient $\omega$ varies based on the strange quark's mass. Specifically, for radiation where mass equals zero, $\omega$ is $1/3$. However, in a scenario where mass is $250\, MeV$, $\omega$ adjusts to $0.28$, positioning $B$ within the range of $41.58\, MeV/fm^3$ to $319.13\, MeV/fm^3$. It is important to note that the MIT bag model, initially presented by \cite{Chodos1}, was motivated to understand the binding of strange quark matter. Later, this model was applied in the context of astrophysical objects, as pointed out by Witten \cite{Chodos2}. It is highlighted that if one considers a star made of quark matter and ignores the contributions due to their masses, the relation between pressure and density is $p = \frac{1}{3}(\rho - 4B)$, where the vacuum pressure $B$ on the Bag wall is responsible for stabilizing the confinement of the quarks. The MIT bag model applies to wormholes and helps us understand how exotic matter needed to keep a wormhole open can be confined in a similar way. The exotic matter has unusual properties, such as negative energy density, which are essential for maintaining the stability of the wormhole throat, preventing it from collapsing and allowing it to remain open for traversable. Researchers has delved into WH configurations within the $f(R, T)$ gravity framework, focusing on SQM characterized by both isotropic and anisotropic pressures \cite{Tayde1}. Moreover, discussions on WH formations leveraging the MIT model have extended into the realm of $f(Q,T)$ gravity, unveiling a novel perspective that supports the existence of navigable WHs within this gravitational theory using strange matter, in alignment with both the Strong Energy Condition (SEC) and the Weak Energy Condition (WEC) \cite{Tayde2}.\\
\indent This study explores WH geometries within $f(Q)$ gravity, employing the MIT bag model alongside various Equation of State (EoS) formulations and the generalized cosmic Chaplygin gas (GCCG) under the constant and various redshift functions. The MIT bag model and the GCCG provide frameworks for understanding and generating the exotic matter required for traversable wormholes. The MIT bag model offers insights into how this matter can be confined, while the GCCG provides a realistic source of the negative pressure needed to maintain the wormhole's stability.\\
\indent 
Several candidates have been suggested that contribute to violating the Null Energy Condition (NEC), which is indicative of phantom-type dark energy. This form of energy is characterized by the equation of state (EoS) \( p = \omega \rho \), where \( p \) and \( \rho \) denote the pressure and energy density of the universe, respectively, with the EoS parameter \( \omega \). Current observational data indicates that dynamic dark energy significantly influences the universe's expansion, which grows over time. This growth in dark energy transitions from \( \omega > -1 \) (the quintessence region with less negative pressure) to \( \omega < -1 \) (the phantom region with more negative pressure) \cite{7} facilitating the formation and stability of WHs. Traversable WHs have garnered interest through various phantom-like EoS models \cite{8,9,10}. Lobo constructed WH solutions using phantom-like Chaplygin gas models \cite{lobo}. He identified several WH models and examined their stability using the averaged null energy condition (ANEC). Chakraborty and Bandyopadhyay \cite{chakraborty} extended this research to modified Chaplygin gas. Jamil et al. \cite{jamil} developed WH models under a polytropic EoS, considering both constant and variable redshift functions, and analyzed their properties. In this paper, we follow the methodology of \cite{lobo,chakraborty,jamil} for constructing the WH solutions through the GCCG model. We evaluate the feasibility and stability of these WH solutions.

Our paper is organized as follows: In section \ref{sec2}, we introduce the basic formalism of $f(Q)$ gravity and discuss the corresponding wormhole field equations in section \ref{sec3}. Section \ref{sec4} provides an overview of the derivation of the equation of motion for the MIT bag model. We then proceed to obtain wormhole solutions for isotropic and anisotropic pressure cases with SQM under the non-constant redshift function in Section \ref{sec5}. Subsequently, we discuss wormhole solutions for the phantom-like GCCG model under different forms of redshift functions in section \ref{sec6}. Additionally, we study the stability analysis and VIQ for both the MIT bag model and GCCG cases in sections \ref{sec7} and \ref{sec8}, respectively. Finally, we conclude our findings in the last section.

\section{$f(Q)$ gravity}\label{sec2}
In the current investigation, we consider that WH exists on the variational Lorentzian spacetime $\mathcal{M}$ which can be explained by the metric tensor $g_{\mu\nu}$, it's determinant $g$ and affine connection $\Gamma$ :
\begin{equation}
    g=g_{\mu\nu}dx^\mu\otimes dx^\nu.
\end{equation}
$\Gamma^\alpha_{\,\,\, \beta}$ is known as the connection of one form, which can be represented in terms of one form of Levi-Civita connection, disformation and contortion tensor \cite{ortin}:
\begin{equation}
\Gamma^{\alpha}_{\,\,\,\,\beta}=w^{\alpha}_{\,\,\,\, \beta}+K^{\alpha}_{\,\,\,\, \beta}+L^{\alpha}_{\,\,\,\,\beta}.
\end{equation}
The above expression can be written as follows:
\begin{equation}
\Gamma^{\alpha}_{\,\,\,\,\, \mu\nu}=\gamma^{\alpha}_{\,\,\,\,\mu\nu}+K^{\alpha}_{\,\,\,\,\mu\nu}+L^{\alpha}_{\,\,\,\,\mu\nu},
    \label{eq:2}
\end{equation}
where $\gamma$, $K$, and $L$ are known as Levi-Civita metric-compatible affine connection, contorsion, and disformation tensors, respectively. As we utilize the $f(Q)$ gravity theory, the symmetric teleparallelism that results from the non-metricity one form and associated tensor properly describes the gravitational sector:
\begin{equation}
    Q^\alpha_{\,\,\,\,\beta}=\Gamma_{(ab)},\quad Q_{\alpha\mu\nu}=\nabla_\alpha g_{\mu\nu},
\end{equation}
and the symmetric portion of the tensor has the following definition:
\begin{equation}
    F_{(\mu\nu)}=\frac{1}{2}\bigg(F_{\mu\nu}+F_{\nu\mu}\bigg).
\end{equation}
When contortion disappears, everything that is left is the disformation tensor.
\begin{equation}\label{b6}
    Q_{\alpha\mu\nu}=-L^{\beta}_{\,\,\,\,\alpha\mu}g_{\beta\nu}-L^{\beta}_{\,\,\,\,\alpha\nu}g_{\beta\mu},
\end{equation}
where disformation tensor is
\begin{equation}
    L^\alpha_{\,\,\,\,\mu\nu}=\frac{1}{2}Q^{\alpha}_{\,\,\,\,\mu\nu}-Q_{(\mu\nu)}^{\,\,\,\,\,\alpha}.
    \label{eq:77}
\end{equation}
The so-called superpotential and non-metricity together make up the non-metricity scalar, which is the main quantity of the gravitational sector in the STEGR formulation.
\begin{equation}\label{b8}
    Q=-P^{\alpha\mu\nu}Q_{\alpha\mu\nu},
\end{equation}
where superpotential has the following complex form:
\begin{equation}
    P^{\alpha}_{\,\,\,\,\mu\nu}=\frac{1}{4}\bigg[2Q^{\alpha}_{\,\,\,\,(\mu\nu)}-Q^{\alpha}_{\,\,\,\,\mu\nu}+Q^\alpha g_{\mu\nu}-\delta^{\alpha}_{(i}Q_{j)}-\overline{Q}^\alpha g_{\mu\nu}\bigg].
\end{equation}
Here, $Q^{\alpha}=Q^\nu_{\,\,\,\,\alpha\nu}$ and $\overline{Q}_\alpha=Q^\mu_{\,\,\,\,\alpha\mu}$ are two independent traces of the non-metricity tensor $Q_{\alpha\mu\nu}=\nabla_\alpha g_{\mu\nu}$. Finally, since we have already supplied all of the essential STEGR formalism terms, we may go on to provide the modified action integral for $f(Q)$ gravity  \cite{xu}:
\begin{equation}
    \mathcal{S}[g,\Gamma,\Psi_i]=\int d^4x \sqrt{-g}f(Q)+\mathcal{S}_{\mathrm{M}}[g,\Gamma,\Psi_i].
    \label{eq:10}
\end{equation}
where, $\mathcal{S}_{\mathrm{M}}[g,\Gamma,\Psi_i]$ presents the action integral which describes the contribution of additional matter field $\Psi_i$ minimally and non-minimally coupled to gravity to the total Einstein-Hilbert action integral. If we will vary equation (\ref{eq:10}) with respect to the metric tensor inverse $g^{\mu\nu}$, we will get the abstract field equations for the theory:
\begin{multline}
    \frac{2}{\sqrt{-g}}\nabla_\gamma(\sqrt{-g}\,f_Q\,P^\gamma\;_{\mu\nu})+\frac{1}{2}g_{\mu\nu}f \\
+f_Q(P_{\mu\gamma i}\,Q_\nu\;^{\gamma i}-2\,Q_{\gamma i \mu}\,P^{\gamma i}\;_\nu)=-T_{\mu\nu},
\label{eq:11}
\end{multline}

Here, $f_Q$ is the derivative of $f$ with respect to the non-metricity $Q$, and $T_{\mu\nu}$ is the energy-momentum tensor for describing the fluid, which can be defined as:
\begin{equation}
    T_{\mu\nu}=-\frac{2}{\sqrt{-g}}\frac{\delta(\sqrt{-g} \mathcal{L}_{\mathrm{M}})}{\delta g^{\mu\nu}}.
\end{equation}
Where $\mathcal{L}_{\mathrm{M}}$ represents the Lagrangian matter density such that $\int d^4x \sqrt{-g} \mathcal{L}_{\mathrm{M}}=\mathcal{S}_{\mathrm{M}}[g,\Gamma,\Psi_i]$. Again by manipulating the action with respect to the affine connection $\Gamma^\alpha_{\,\,\,\,\mu\nu}$ we get:
\begin{equation}
\nabla_\mu \nabla_\nu (\sqrt{-g}\,f_Q\,P^\gamma\;_{\mu\nu})=0.
\label{eq:12}
\end{equation}
One can study this theory using a coincident gauge involving a specific coordinate choice. In this gauge, the connection disappears, and the non-metricity expressed in Eq. \eqref{b6} can be simplified to the form
\begin{equation}\label{2222}
Q_{\alpha\mu\nu}=\partial_\alpha g_{\mu\nu}.
\end{equation}
This simplification makes calculations easier since only the metric is considered a fundamental variable. However, it should be noted that the action is no longer diffeomorphism invariant in this case, except for standard GR, as stated in \cite{Koivisto}. By employing this coincident gauge, researchers has explored novel solutions for $f(Q)$ theory in the context of black holes, regular black holes, and black-bounce spacetime \cite{rev1}. By choosing the coincident gauge, the degrees of freedom associated with the connection are eliminated. Moreover, in this special choice of coordinate transform, the formalism of $f(Q)$ gravity can be more directly compared with GR. This comparison can help identify deviations from GR and understand the modifications introduced by the $f(Q)$ framework. Moreover,
for numerical simulations, the coincident gauge can be particularly advantageous as it reduces the computational resources needed by simplifying the equations and reducing the number of variables to be evolved.

\section{WH model in $f(Q)$ gravity}
\label{sec3}

In this section, our goal is to outline the basic structure of WH theory and briefly summarize the $f(Q)$ gravity approach. We focus on the spherically symmetric static Morris-Thorne WH metric \cite{Morris}, which is characterized by the following equation:
\begin{equation}\label{c1}
ds^2=-e^{2\phi(r)}dt^2+\frac{dr^2}{1-\frac{b(r)}{r}}+r^2(d\theta^2+\sin^2\theta d\Phi^2),
\end{equation}
where the functions $b(r)$ and $\phi(r)$ are identified as the shape and redshift functions, respectively. A crucial aspect for a WH to be considered traversable is the flaring-out condition, which mathematically can be represented as:
\[\frac{b - b'r}{b^2} > 0.\]
At the throat of the WH, this condition simplifies to:
\[b'(r_0) < 1.\]
Moreover, for any location beyond the throat, that is, when $r > r_0$, it is necessary that $1 - \frac{b(r)}{r} > 0$ is satisfied. It is also essential for the WH to be asymptotically flat, which implies:
\[\frac{b(r)}{r} \to 0 \text{ as } r \to \infty.\]
Finally, the redshift function must remain finite across the entire structure to avoid an event horizon.\\
Also, another significant criterion is the proper radial distance $l(r)$, represented as
\begin{equation}
l(r)=\pm \int_{r_0}^{r}\frac{dr}{\sqrt{1-\frac{b(r)}{r}}},
\end{equation}
is needed to be finite everywhere. Here, the $\pm$ symbols indicate the upper and lower portions of the wormhole, which are linked by the throat. Also, the proper distance decreases from the upper universe $l=+\infty$ to
the throat and then from  $l=0$ to $-\infty$ in the lower universe. Moreover, $l$ should be greater than or equal to the coordinate distance $\mid l(r)\mid \geq r-r_0$. The embedding surface of the wormhole can be obeyed by defining the embedding surface $z(r)$ at a fixed $\theta=\pi/2$ and time $t=\text{constant}$. Hence, equation \eqref{c1} reduces to
\begin{equation}
\label{6a}
ds^2=\left(1-\frac{b(r)}{r}\right)^{-1}dr^2+r^2 d\Phi^2.
\end{equation}
The above metric can be embedded into three-dimensional Euclidean space with cylindrical coordinates $r,\,\ \phi$ and $z$ as
\begin{equation}
\label{6b}
ds^2=dz^2+dr^2+r^2 d\Phi^2.
\end{equation}
Now, on comparing equations \eqref{6a} and \eqref{6b}, we obtained the following slope equation so that by integrating it, one can find the embedding surface $z(r)$
\begin{equation}
\label{6c}
\frac{dz}{dr}=\pm \sqrt{\frac{r}{r-b(r)}-1}.
\end{equation}

Now for the given metric \eqref{c1}, the non-metricity scalar, derived from equation \eqref{b8}, is represented as follows:
\begin{equation}
Q = -\frac{2}{r} \left(1 - \frac{b}{r}\right) \left(2\phi' + \frac{1}{r}\right).
\end{equation}
Furthermore, the consideration is on matter characterized by an anisotropic fluid, which is expressed in the form
\begin{equation}
T^{\mu}_{\nu} = \text{diag}[-\rho, p_r, p_t, p_t],
\end{equation}
where $\rho$ is the energy density, and $p_r$ and $p_t$ represent the radial and tangential pressures, respectively.\\
Now, the field equations for the metric \eqref{c1} under anisotropic fluid within the context of the modified symmetric teleparallel gravity theory can be obtained as
\begin{multline}
\label{eq:16}
\left[\frac{1}{r}\left(-\frac{1}{r}+\frac{rb^{'}+b}{r^2}-2\phi^{'}\left(1-\frac{b}{r}\right)\right)\right]f_Q \\
-\frac{2}{r}\left(1-\frac{b}{r}\right)f_{QQ}Q^{'}-\frac{f}{2}=-\rho,
\end{multline}
\begin{equation}
\label{eq:17}
\left[\frac{2}{r}\left(1-\frac{b}{r}\right)\left(2\phi^{'}+\frac{1}{r}\right)-\frac{1}{r^2}\right]f_Q+\frac{f}{2}=-p_r,
\end{equation}
\begin{multline}
\label{eq:18}
\left[\frac{1}{r}\left(\left(1-\frac{b}{r}\right)\left(\frac{1}{r}+\phi^{'}\left(3+r\phi^{'}\right)+r\phi^{''}\right)-\frac{rb^{'}-b}{2r^2}\right.\right.\\\left.\left.
\left(1+r\phi^{'}\right)\right)\right]f_Q+\frac{1}{r}\left(1-\frac{b}{r}\right)\left(1+r\phi^{'}\right)f_{QQ}Q^{'}\\
+\frac{f}{2}=-p_t,
\end{multline}
\begin{equation}\label{111}
\frac{\cot{\theta}}{2}f_{QQ}Q^{'}=0,
\end{equation}
where ${'}$ represents $\frac{d}{dr}$. Now, looking at the off-diagonal component given in Eq. \eqref{111} gives the following relation
\begin{equation}\label{1111}
f_{QQ}=0 \implies f(Q)=\alpha Q+ \beta,
\end{equation}
where, $\alpha$ and $\beta$ are constants.

For the above model \eqref{1111}, the field equations (\ref{eq:16}-\ref{eq:18}) turn out to be
\begin{eqnarray} \label{fe1}
   &&\hspace{1cm} \rho = \frac{\beta }{2}-\frac{\alpha~ b^{\prime}}{r^2},\\ \label{fe2}
    &&\hspace{1cm} p_r=\frac{\alpha  (2 r (b-r) \phi '+b}{r^3}-\frac{\beta }{2},\\ \label{fe3}
    &&\hspace{1cm} p_t= \frac{\alpha  (r \phi '+1) (r b'+2 r (b-r) \phi '-b}{2 r^3}\nonumber\\&&\hspace{2cm}+\frac{\alpha  (b-r) \phi ''}{r}-\frac{\beta }{2}.
\end{eqnarray}
It can be noted that the above model can be reduced to GR if we consider $\alpha=-1$ and $\beta= 0$.
\begin{itemize}
    \item \textbf{Energy conditions :} \\
the energy conditions, which are derived from Raychaudhuri equations, establish inequalities among thermodynamic parameters such as $\rho, p_r$, and $p_t$ \cite{RR}. These conditions illustrate the behavior of congruences and the gravitational attraction for timelike, spacelike, or lightlike curves. Primarily, energy conditions serve as essential tools to characterize strong gravitational fields, as they accurately describe the geodesic structure of spacetime. Additionally, these conditions help assess the realism of matter distribution and the potential existence of WH geometries. Given that our analysis involves anisotropic fluid, the relevant energy conditions are as follows:
\end{itemize}
    
\begin{enumerate}
    \item \textit{NEC (Null Energy Condition)}:
    \begin{enumerate}
        \item Physical form : For any null vector $l^{m}$, $T_{mn}l^{m}l^{n}\geq 0$
        \item Effective form : $\rho+p_r\geq0$, $\rho+p_t\geq0$.
    \end{enumerate}
    \item \textit{WEC (Weak Energy Condition)}:
    \begin{enumerate}
        \item Physical form : For any timelike vector $t^{m}$, $T_{mn}t^{m}t^{n}\geq 0$
        \item Effective form : $\rho \geq 0$, $\rho+p_r\geq0$, $\rho+p_t\geq0$.
    \end{enumerate}
    \item \textit{SEC (Strong Energy Condition)}:
    \begin{enumerate}
        \item Physical form : For any timelike vector $t^{m}$, $(T_{mn}-\frac{1}{2}T g_{mn})t^{m}t^{n}\geq 0$
        \item Effective form : $\rho \geq 0$, $\rho+p_r\geq0$, $\rho+p_t\geq0$, $\rho+p_r+2p_t\geq0$.
    \end{enumerate}
    \item \textit{DEC (Dominant Energy Condition)}:
    \begin{enumerate}
        \item Physical form: For any two co-oriented times like vectors  $t^{m}$, and $\eta^{n}$ $T_{mn}t^{m}\eta^{n}\geq 0$
        \item Effective form : $\rho \geq 0$, $\rho\pm p_r\geq0$, $\rho\pm p_t\geq0$.
    \end{enumerate}

\begin{center}
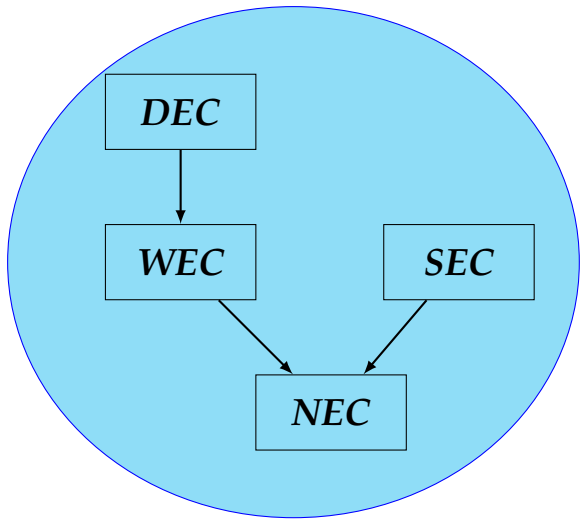

\begin{tikzpicture}
\draw[blue,fill= bluE] (7,0) ellipse (3.8 and 3.4);
\node[rectangle,draw = black, minimum width = 2cm, minimum height = 1cm] (dec) at (5.5,2) {\Large \textit{\textbf{DEC}}};
\node[rectangle,draw = black, minimum width = 2cm, minimum height = 1cm] (wec) at (5.5,0) {\Large \textit{\textbf{WEC}}};
\node[rectangle,draw = black, minimum width = 2cm, minimum height = 1cm] (nec) at (7.5,-2) {\Large \textit{\textbf{NEC}}};
\node[rectangle,draw = black, minimum width = 2cm, minimum height = 1cm] (sec) at (9.2,0) {\Large \textit{\textbf{SEC}}};
\draw[-latex, thick] (dec) to (wec);
\draw[-latex, thick] (wec) to (nec);
\draw[-latex, thick] (sec) to (nec);
\end{tikzpicture}
\captionof{figure}{A schematic diagram among the relation of energy condition: An arrow from A to B in the illustration indicates that A implies B.}
\label{f1100}
\end{center}
The physical implications of the energy conditions, namely NEC, WEC, SEC, and DEC, can be summarized as follows:
\begin{enumerate}
    \item  \textit{Weak Energy Condition (WEC):} This condition asserts that the energy density ($\rho$) measured by any observer along a timelike vector must always be non-negative, i.e., $\rho \geq 0$.
    \item \textit{Strong Energy Condition (SEC):} The SEC indicates that the gravitational field should always be attractive, suggesting the presence of a strong gravitational field.
    \item \textit{Dominant Energy Condition (DEC):} According to the DEC, the energy flux measured by any observer, whether timelike or null, must always be non-negative, i.e., $\rho \geq 0$.
    \item \textit{Null Energy Condition (NEC):} The NEC serves as a basic requirement for both the SEC and the WEC, implying that $\text{WEC} \subset \text{NEC}$ and $\text{SEC} \subset \text{NEC}$.
\end{enumerate}
\end{enumerate}
Also, we have presented the implications among the energy conditions in Fig. \ref{f1100}.

\section{Thermodynamics of  density-dependent B parameter in MIT Bag Model}
\label{sec4}
The study of matter at extremely high densities is a crucial topic in contemporary physics. This issue is complex not only theoretically but also due to the limitations of laboratory experiments, which fail to provide comprehensive data for a full understanding. To test our grasp of the relevant physics, we must look for astrophysics and the behavior of compact general relativistic objects. Neutron stars, in particular, serve as unique laboratories for such extreme physics. Their core densities can reach levels about ten times higher than nuclear saturation, potentially containing exotic states of matter, such as hyperon phases with net strangeness and deconfined quarks \cite{hit1}. 

Next, let us briefly discuss the background of the MIT Bag model. Witten \cite{wi} initially proposed that strange quark matter (SQM) could represent a genuine ground state of Quantum Chromodynamics (QCD). Subsequently, SQM was investigated as the constituents of a Fermi gas consisting of up, down, and massive strange quarks (denoted by s with mass $m_s$). In the MIT Bag model, where the pressure ($p$) is given by  $p=\omega(\rho-4B)$, the vacuum pressure ($B$) acting on the Bag wall balances the pressure exerted by the quarks, thereby stabilizing the system. The occurrence of the de-confinement phase transition relies on both the temperature and the baryon number density of the system at high densities. The parameter $B$, dependent on density, has been extensively studied in the literature \cite{m3,m4,m5}.

It is well understood that the energy density and pressure of a particle system can be calculated using a general ensemble theory. The physical characteristics of this particle system can be derived from the corresponding partition function \cite{hit2,hit3}.
\begin{equation} \label{8}
\mathcal {Z}=\sum _{N_i, \epsilon } e^{ - \zeta (E_{N_i, \epsilon } - \sum _i N_i \mu _i) } \,,
\end{equation}
where $\zeta = \frac{1}{ k_B T}$, $N_i$ represents the particle number, and $\mu_i$ stands for chemical potential. It is evident that, in general, the microscopic energy \( E_{N_i, \epsilon} \) depends on the number of particles \( N_i \), the masses of the particles \( m_i \), the volume of the system \( V \), and other quantum numbers \( \epsilon \). Thus, we can express this relationship as:

\[ E_{N_i, \epsilon} = f(N_i, m_i, V, \epsilon) \]

The pressure of the system can be obtained from the following relation.

\begin{eqnarray}\label{9}
p_0 = -\Xi = \frac{1}{\zeta V} \ln \mathcal {Z}\,,
\end{eqnarray}
here $\Xi$, stands for the thermodynamic potential, that depends on the chemical potential $\mu_i$, temperature $T$ of the system, and the mass of the particle $m_i$. The statistical average of the energy density and the particle number $N_i$ has the following form respectively,
\begin{eqnarray}\label{11}
&&\hspace{0.0cm}\bar{E} = - \frac{\partial }{\partial \zeta } \ln \mathcal {Z} + \sum _i \bar{N}_i \mu _i\,,\\
\label{12}
&&\hspace{-0.5cm}\bar{N}_i = \frac{1}{\zeta } \left( \frac{\partial }{\partial \mu _i} \ln \mathcal {Z} \right) _{T, V, m_j} =-V \left( \frac{\partial \Xi }{\partial \mu _i} \right) _{T, m_j}\,.
\end{eqnarray}
The total energy of the whole system can be evaluated as:
\begin{eqnarray}\label{13}
E_0 = \Xi + \sum _i n_i \mu _i\,,
\end{eqnarray}
Where particle number density is given by :
\begin{equation}\label{14}
n_i = \frac{\bar{N}_i }{V} =- \left( \frac{\partial \Xi }{\partial \mu _i}\right) _{T, m_j} \,.
\end{equation}
For the MIT bag model, the microscopic energy of the strange matter system is given by
\begin{equation}\label{15}
E_{N_i, \epsilon }^{Bag}= E_{N_i, \epsilon } + B V\,,
\end{equation}
where the form of the correspondent partition function is :
\begin{eqnarray}\label{16}
\mathcal {Z}^{Bag} =\mathcal {Z} e^{- \zeta B V}\,.
\end{eqnarray}
Hence, the particle number density is given by:
\begin{equation}\label{17}
n_i^{Bag} = \frac{\bar{N}_i}{V} =  - \left( \frac{\partial }{\partial \mu _i} (\Xi +B) \right) _{T, m_j, E_{N_i, \epsilon }, B }\,,
\end{equation}
where the energy and the bag pressure can be written as :
\begin{eqnarray}\label{18}
p^{B a g} & =\frac{1}{\zeta} \frac{\partial}{\partial V}\left(\ln \mathcal{Z}^{B a g}\right)=\frac{1}{\zeta} \frac{\partial}{\partial V}(\ln \mathcal{Z}-\zeta B V) \nonumber\\
& =-(\Xi+B)-V \frac{\partial}{\partial V}(\Xi+B) \\
E^{B a g} & =(\Xi+B)+\sum_{i} n_{i} \mu_{i}-T \frac{\partial}{\partial T}(\Xi+B) 
\end{eqnarray}
If the particle mass is independent of the baryon density and the Bag parameter $B$ depends on system density, then the above three equations are represented as,
\begin{eqnarray}
 &&\hspace{1cm}n_{i}^{B a g}=-\left(\frac{\partial \Xi}{\partial \mu_{i}}\right) \\
 &&\hspace{1cm}p^{B a g}=-(\Xi+B)+n_{b} V \frac{\partial B}{\partial n_{b}} \\
&&\hspace{1cm}E^{B a g}=(\Xi+B)+\sum_{i} n_{i} \mu_{i} 
\end{eqnarray}

Demonstrating the equation of state for the MIT bag model, our next section aims to identify wormhole families that satisfy the energy conditions within 
$f(Q)$ gravity. These wormholes will have density and radial pressure that adhere to the MIT bag model equation of state. Additionally, we plan to explore the characteristics of a bag parameter that varies with the radial coordinate.

\section{Role of density-dependent Bag parameter of MIT bag model in WH model}
\label{sec5}
\subsection{Model-I: Isotropic fluid}
For isotropic fluid distribution, ($p_r=p_t=p$) the above three field equations (\ref{fe1}-\ref{fe3}) consists of four unknown physical quantities $\small\{\rho(r),p(r),b(r),\phi(r)\small\}$. To address this system, various strategies have been employed in the literature. One approach involves modeling an appropriate spacetime geometry by selecting a specific equation of state and defining either $\phi(r)$ or $b(r)$, thereby resolving the system of coupled differential equations. Another method involves specifying the forms of the functions $b(r)$ and $\phi(r)$ directly and subsequently determining the stress-energy tensor components. Alternatively, one could define a suitable source for the spacetime geometry by imposing the stress-energy components and then determining the metric fields. For that, we are considering a variable redshift function $\phi(r)=\text{Log}\left(1+\frac{r_0}{r}\right)$. Using the above redshift function, Scientists Pavlovic and Sossich \cite{phi1} studied potential WH solutions within the framework of four distinct $f (R)$ models: the MJWQ model \cite{1}, the exponential model \cite{2,3}, the Tsujikawa model \cite{4,5}, and the Starobinsky model \cite{6,71}. They introduced the redshift function $\phi(r) = \text{Log}(1+\frac{r_0}{r})$ and successfully derived WH solutions for the first three models, notably without requiring exotic matter. This outcome holds substantial significance. This suggests that the choice of redshift function may play a crucial role. Their findings prompted exploring the validity of energy conditions in other $f (R)$ models and the type of matter necessary to uphold WH solutions. 
Hence, motivated by their work, in our current research, we adopt the same redshift function: $\phi(r) = \text{Log}(1+\frac{r_0}{r})$ to assess energy conditions and investigate WH solutions for quark matter supported WH. Now, using the isotropy criteria and the known redshift function from the Eqs. \eqref{fe1} and \eqref{fe2}, we have determined the shape function $b(r)$  as
\begin{eqnarray}\label{d1}
b(r)=3r\Big[\frac{1}{2}+\frac{r}{4r_0}+\big(\frac{r}{4r_0}\big)^2 \Big]+C_1 r^3 e^{\frac{4 r_0}{r}}.
\end{eqnarray}
where $C_1$ is the integrating constant. Now, to find the value of $C_1$, we impose throat condition $b(r_0)=r_0$ to above expression, we obtain
\begin{equation}
C_1=-\frac{23}{16 e^4 r_0^2}.
\end{equation}
After substituting the values of $C_1$ to the Eq. \eqref{d1}, one can obtain the final shape function
\begin{equation}
 b(r)= \frac{r \Big[r^2 \big(3-23 e^{\frac{4 r_0}{r}-4}\big)+12 r r_0+24 r_0^2\Big]}{16 r_0^2}
\end{equation}
 It is clear that the shape function does not follow the asymptotic condition, i.e.
 \begin{equation}
    \lim_{r\rightarrow \infty} \frac{b(r)}{r} \nrightarrow 0.
 \end{equation}

\begin{figure}
   \includegraphics[width=7.cm, height=4.5cm]{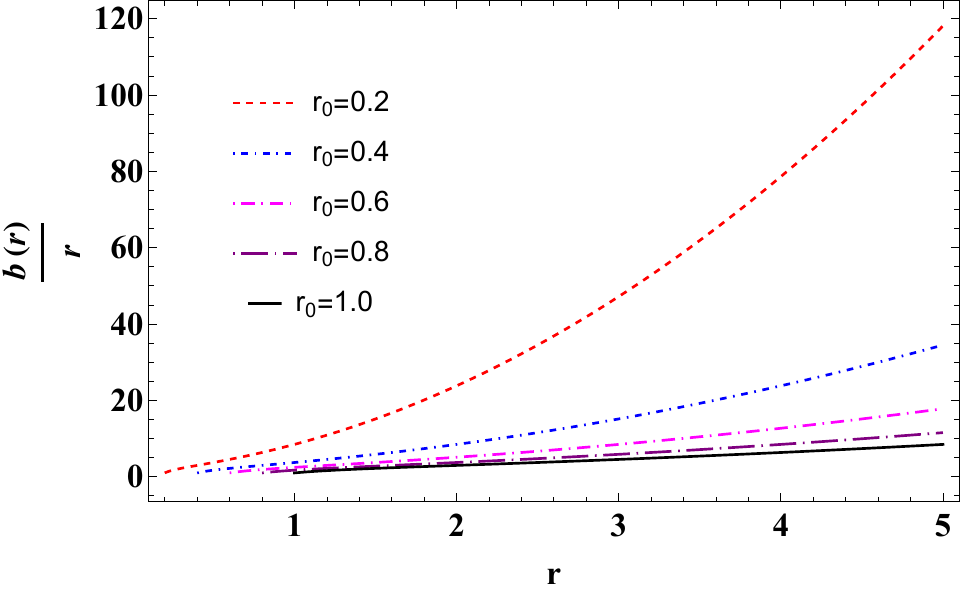}~
      \caption{Model-I: Violation of asymptotic flatness condition :} \label{f111}  
\end{figure}
 
For more clarification, we have shown it graphically in Fig. \ref{f111}. One can see that for any value of throat radius, $r_0\in[0.2,1.0]$, $ \frac{b(r)}{r} \nrightarrow 0$ as $r\to \infty$.
This result is highly significant, indicating that an isotropic WH is not theoretically possible. similar results can be seen in GR \cite{Liempi} as well as in $f(R,L_m)$ gravity \cite{Solanki}. Thus, it is safe to conclude that asymptotic wormhole solutions under isotropic pressure may not be possible.

 So, for the known redshift function $\phi(r)$ and shape function $b(r)$ we get the density ($\rho$) and pressure ($p$) of the fluid as,
\begin{eqnarray}
   &&\hspace{0cm} \rho=\frac{1}{16 e^4 r^2 r_0^2}\Big[e^4 \left(8 \beta  r^2 r_0^2-3 \alpha  \left(3 r^2+8 r r_0+8 r_0^2\right)\right)\\ \nonumber&&\hspace{2cm}+23 \alpha  r e^{\frac{4 r_0}{r}} (3 r-4 r_0)\Big],\\
  &&\hspace{0cm} p = \frac{1}{16 e^4 r^2 r_0^2 (r+r_0)}\Big[e^4 \big(\alpha  \big(3 r^3+9 r^2 r_0+12 r r_0^2\\\nonumber&&\hspace{1cm}+8 r_0^3\big)-8 \beta  r^2 r_0^2 (r+r_0)\big)-23 \alpha  r^2 e^{\frac{4 r_0}{r}} (r-r_0)\Big].
\end{eqnarray}
 In this current article, we are interested in exploring the role of density-dependent Bag parameter $B$ for our constructed WH. Therefore by using the MIT Bag EoS, $p=\omega(\rho-4B)$ we have determined the Bag parameter expression in terms of radial coordinate $r$ given by:
\begin{eqnarray}                                
     &&\hspace{0cm}B(r)=\frac{(r+r_0)^{-1}}{64e^4 r^2r_0^2 \omega  }\big\{-9 e^4 \alpha  r^3 \omega -3 e^4 \alpha  r^3+8 e^4 \beta  r^3 r_0^2 \omega  \nonumber\\ &&\hspace{1cm} +8 e^4 \beta  r^3 r_0^2+69 \alpha  r^3 \omega  e^{\frac{4 r_0}{r}}+23 \alpha  r^3 e^{\frac{4 r_0}{r}}+8 e^4 \beta  r^2 r_0^3 \omega  \nonumber \\ &&\hspace{1cm}+8 e^4 \beta  r^2 r_0^3-23 \alpha  r^2 r_0 \omega  e^{\frac{4 r_0}{r}}-33 e^4 \alpha  r^2 r_0 \omega -8 e^4 \alpha  r_0^3 \nonumber \\ &&\hspace{1cm}-23 \alpha  r^2 r_0 e^{\frac{4 r_0}{r}}-92 \alpha  r r_0^2 \omega  e^{\frac{4 r_0}{r}}-48 e^4 \alpha  r r_0^2 \omega \nonumber\\ &&\hspace{1cm} -12 e^4 \alpha  r r_0^2-24 e^4 \alpha  r_0^3 \omega -9 e^4 \alpha  r^2 r_0\big\}.\label{br12}
\end{eqnarray}

\begin{figure*}
   \includegraphics[width=5.5cm, height=4.1cm]{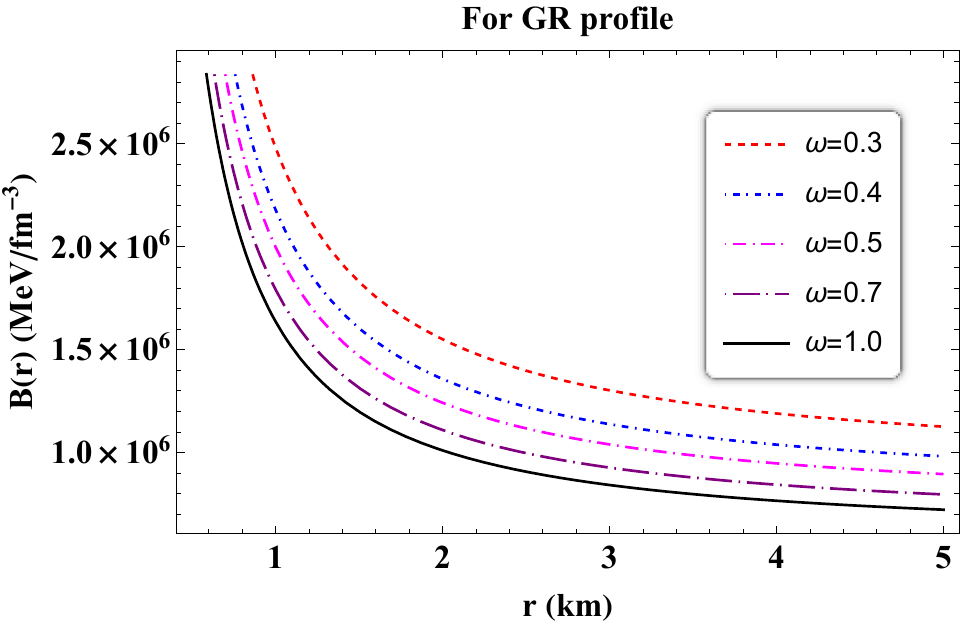}~
    \includegraphics[width=5.5cm, height=4.1cm]{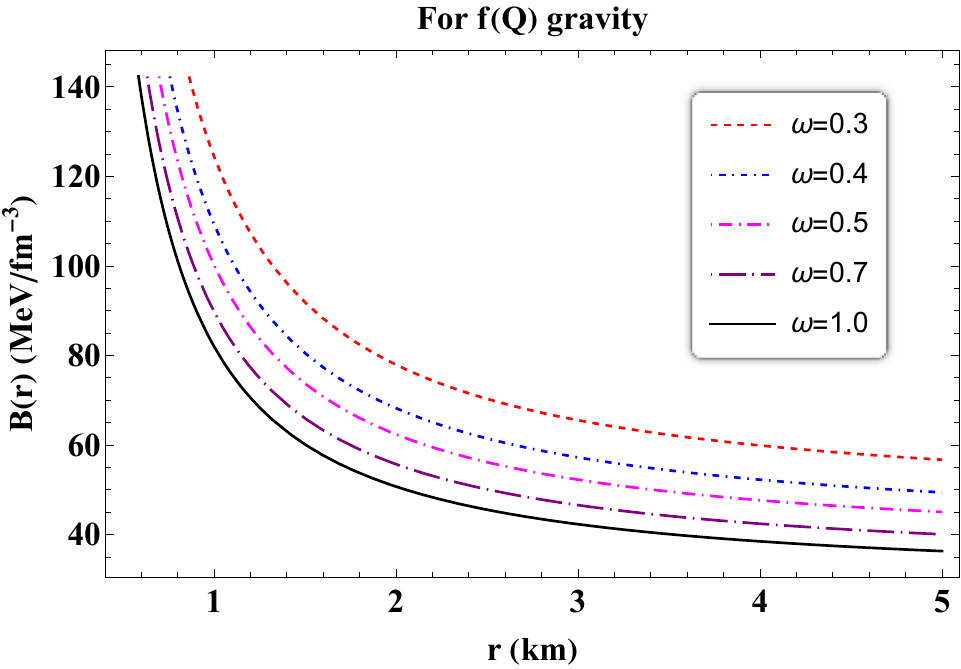}~
    \includegraphics[width=6.0cm, height=4.3cm]{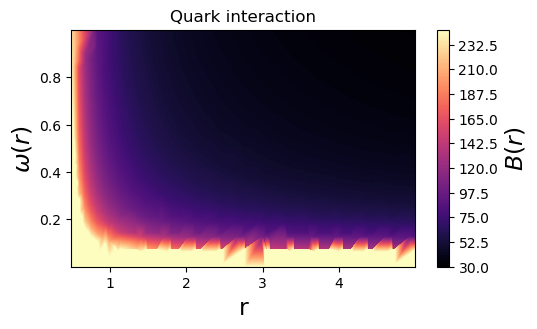}
    
      \caption{\textbf{Isotropic Model}: Left panel : Profile of $B(r)$ for GR varying $\omega$, ($\alpha=-1$, and $\beta=0$). Middle panel: Profile of $B(r)$ for $f(Q)$ gravity varying $\omega$, ($\alpha=-0.0005$ and $\beta=0.0001$), Right panel : Contour plot of $B(r)$ w.r.t. $r$ and $\omega\in[0,1]$ ($\alpha=-0.0005$ and $\beta=0.0001$).} \label{f1}  
\end{figure*}

    

From the above expression (\ref{br12}), it is notable that, $B(r_0)=\frac{8 \beta  r_0^2 (\omega +1)-\alpha  (57 \omega +16)}{64 r_0^2 \omega }$ is a constant at the WH throat $r = r_0$
for a set of values of $\omega,\, \alpha, \text{and} \,\beta$. Within the specified range of $41.58 MeV/fm^3 < B < 319.13 MeV/fm^3$, a preferred value for the bag parameter can be established at the throat, varying based on the assigned set parameters \cite{mit1,mit2}. However, as one moves away from the throat, the value of $B(r)$ diminishes rapidly, ultimately approaching zero at significantly distant points. We have graphically analyzed the behavior of $B(r)$ in Fig. \ref{f1}. In the initial two panels of the figure, we conducted a comparative analysis of $B(r)$ between General Relativity (GR) and the $f(Q)$ model. It's apparent that in the $f(Q)$ model, at the WH throat where $r=r_0$, it possesses a value of $140\, \text{MeV/fm}^3$, falling within the range of $[41.58, 319.13]\, \text{MeV/fm}^3$ by varying a wide range of $\omega\in[0.3,1.0]$. However, in the case of GR, the value is on the order of $10^6$, which significantly exceeds the relevant range of $[41.58, 319.13]\, \text{MeV/fm}^3$, rendering it inconsequential. In the third panel of Fig. \ref{f1}, we present a 3D contour plot of $B(r)\text{MeV/fm}^3$, where both $r$ and $\omega(r)$ are varied. It is evident that as $\omega$ increases, the value of $B$ decreases at the throat. However, it tends towards zero as one moves away from the throat. The bag parameter $B(r)$ plays a crucial role in the model and affects the overall behavior of quark matter. These connections help us understand how exotic forms of matter (such as quark matter) might influence the overall cosmic dynamics. Fig. \ref{f1} illustrates that as $\omega\to 0$, the bag pressure $B(r)$ increases and approaches its upper limit. This indicates that during the thermodynamic phase transition from the matter-dominated to the radiation-dominated era, our model suggests enhanced quark interactions within the WH. However, given that the asymptotic flatness condition doesn't hold in the isotropic WH model, we aim to explore the anisotropic quark matter-supported WH model in the next subsection.

\subsection{Model-II: Anisotropic Fluid}
Here, in this section, we study the geometry of a WH spacetime filled with anisotropic matter distribution where $p_r\neq p_t$. In this case, the MIT bag model can be represented as
\begin{equation}\label{4b1}
p_r=\omega(\rho-4B).
\end{equation}
Now, For the known redshift function $\phi(r)=Log(1+\frac{r_0}{r})$ and using the Eqs. (\ref{fe1}) and (\ref{fe2}) in the above MIT bag model EoS \eqref{4b1}, we obtain the bag parameter $B(r)$ in the form :
\begin{eqnarray}\label{bag1}
    B(r)=\frac{\beta}{8}+\frac{\beta}{8\omega}-\frac{\alpha  r_0}{2 r^2 \omega  (r+r_0)}+\frac{\alpha  b(r) (r_0-r)}{4 r^3 \omega  (r+r_0)} -\frac{\alpha  b'(r)}{4 r^2}~~~~~~~~~
\end{eqnarray}
Here, $B(r)$ can be determined for a given shape function. We shall consider shape function $b(r)$ that is different from that obtained in the isotropic case. In the next segment, we explore two particular shape functions for constructing the WH model.

\subsubsection{Shape function: I}
Throughout this section, we examine WHs characterized by a shape function tailored to ensure compatibility with an asymptotically flat regime. Specifically, we focus on WHs featuring the following shape function \cite{fsn}
\begin{eqnarray}\label{shf1}
    b(r)=r_0+ar_0\Big[\big(\frac{r}{r_0}\big)^{b_1}-1\Big].
\end{eqnarray}
Where $a$ and $b_1$ are the arbitrary constants. The requirement $a b_1>0$ becomes necessary in this specific shape function for considering a positive energy density. To ensure satisfaction of the flaring-out condition at the throat ($r_0$), 
 leading to the additional constraint $a b_1<1$. In summary, the parameters are subject to the following restrictions.
 \begin{eqnarray}
     0 < a b_1 < 1, \quad b_1 <1.
 \end{eqnarray}
By using this shape function (\ref{shf1}), in Eq. (\ref{bag1}), we have determined the density-dependent bag parameter expression as
\begin{eqnarray}
    B(r)=\frac{1}{8 r^3 \omega  (r+r_0)}\Big[\big\{-2 a \alpha  b_1 r_0^2 \omega  +2 a \alpha  r_0^2 -2 a \alpha  b_1 r r_0 \omega \nonumber\\&&\hspace{-7.2cm} -2 a \alpha  r r_0 \big\}(\frac{r}{r_0})^{b_1}+2 a \alpha  r r_0-2 a \alpha  r_0^2+\beta  r^4 \omega \nonumber\\&&\hspace{-7.2cm}+\beta  r^4+\beta  r^3 r_0 \omega +\beta  r^3 r_0-6 \alpha  r r_0+2 \alpha  r_0^2\Big].
\end{eqnarray}

\begin{figure*}
   \includegraphics[width=5.5cm, height=4.1cm]{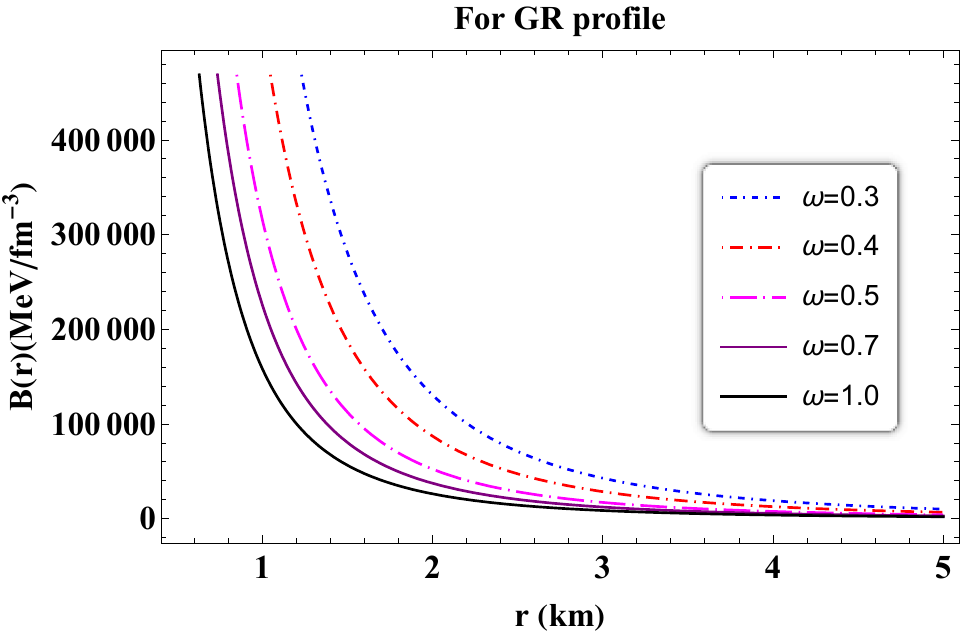}~
    \includegraphics[width=5.5cm, height=4.1cm]{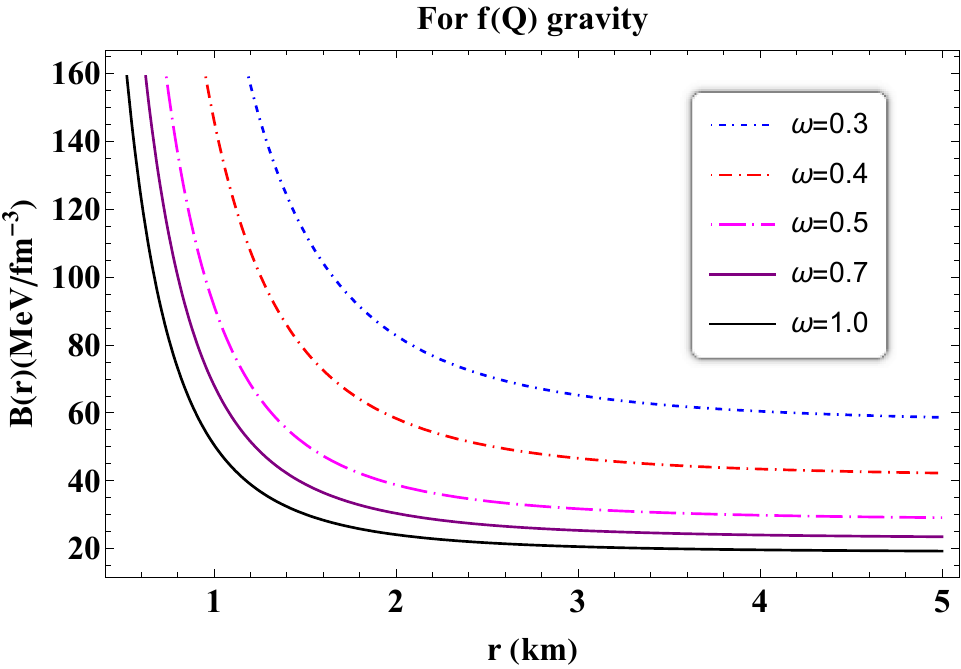}~
    \includegraphics[width=6.0cm, height=4.2cm]{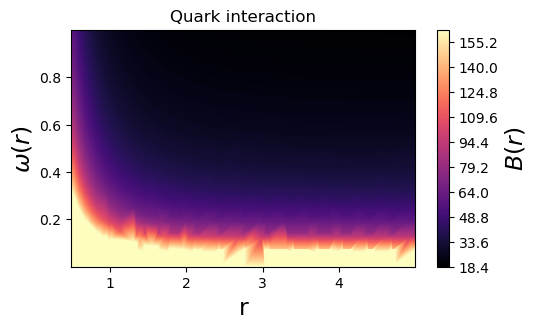}
      \caption{(\textbf{Anisotropic Model, shape function I}): Left panel : Profile of $B(r)$ w.r.t. $r$ for GR ($\alpha=-1$, $\beta=0$, $a=0.2$, $b_1=0.5$). Middle panel: Profile of $B(r)$ w.r.t. $r$ for $f(Q)$ gravity ($\alpha=-0.0005$, $\beta=0.0001$, $a=0.2$, $b_1=0.5$). Right panel : Contour plot of $B(r)$ w.r.t. $r$ and $\omega\in[0,1]$ ($\alpha=-0.0005$ and $\beta=0.0001$, $a=0.2$, $b_1=0.5$).} \label{f2}  
\end{figure*}

\begin{figure*}
   \includegraphics[width=6.2cm, height=4.1cm]{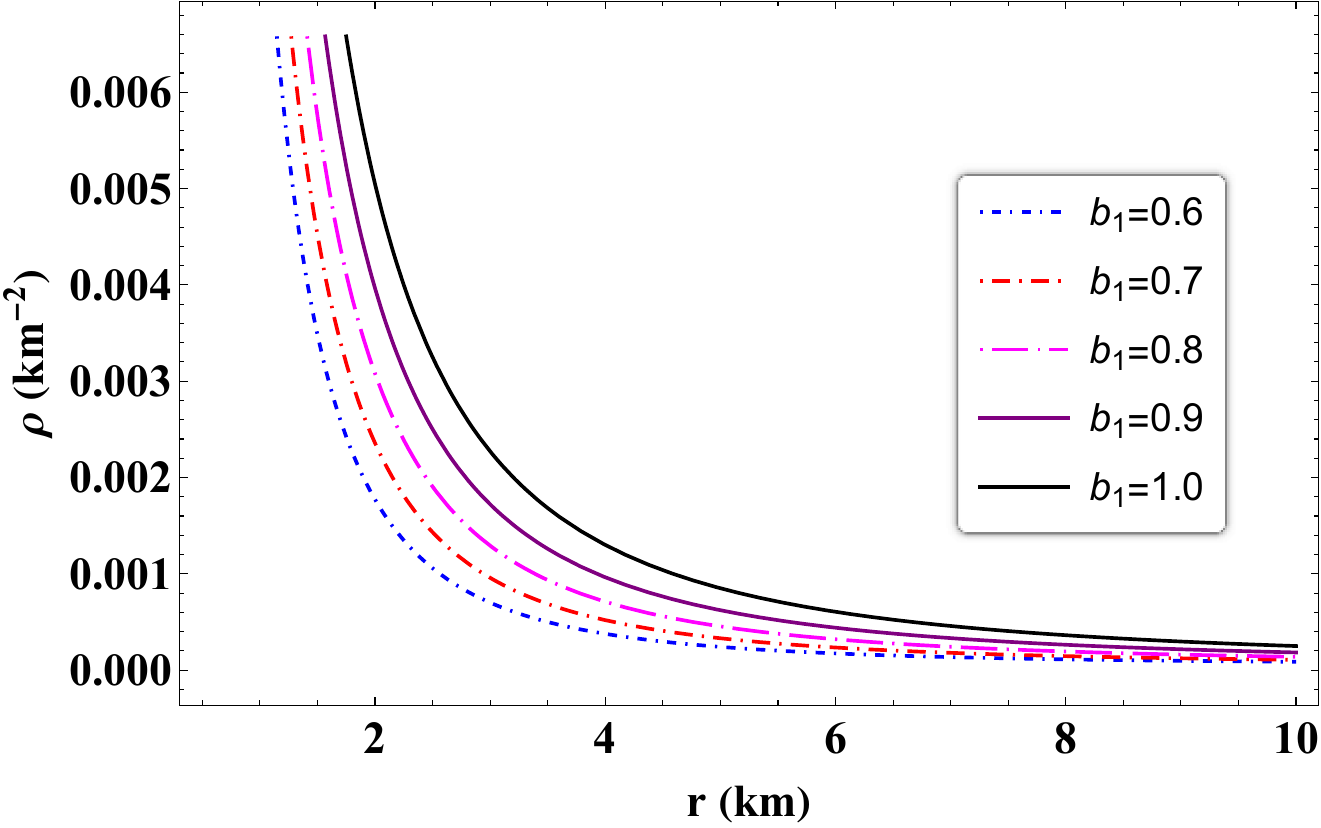}~
    \includegraphics[width=6.2cm, height=4.1cm]{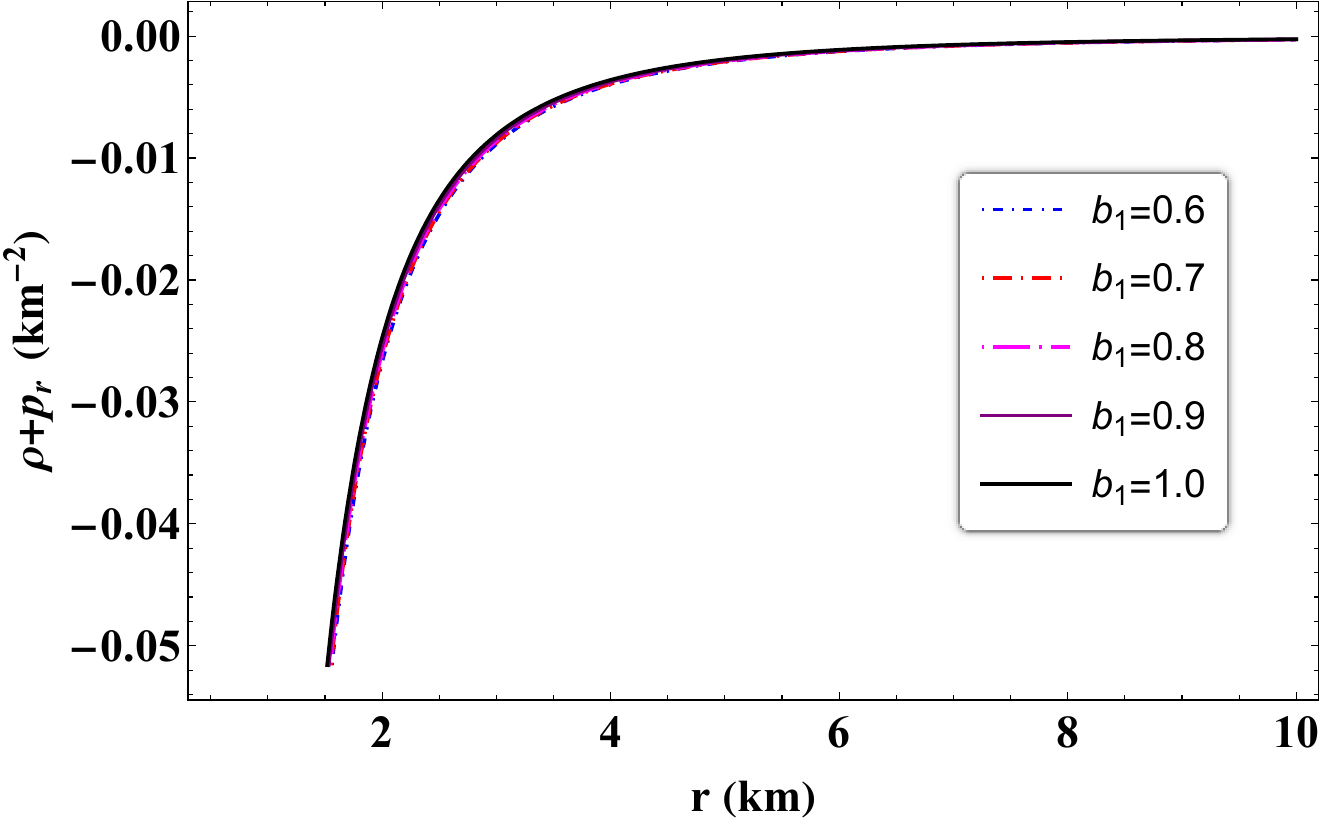}~
    \includegraphics[width=6.2cm, height=4.1cm]{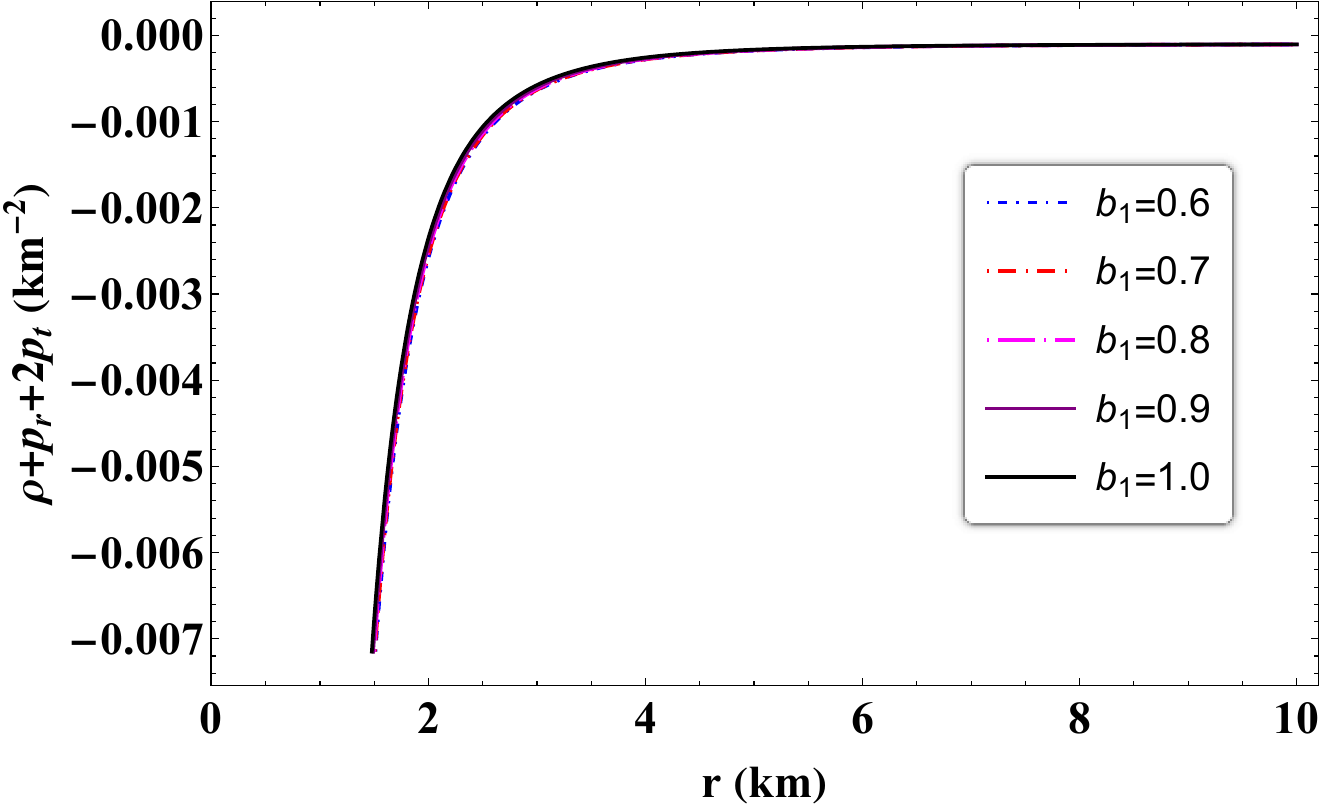}
      \caption{(\textbf{Model-II, shape function I}) Profile of various energy conditions w.r.t. $'r'$ for $\alpha=-0.0005$ and $\beta=0.0001$, $a=0.2$,.} \label{f3}  
\end{figure*}

\textbf{Physical analysis of $B(r)$:} We have analyzed graphically the profile of density-dependent Bag parameter $B(r)$ in Fig. \ref{f2}. We have done a comparative analysis between GR and $f(Q)$ gravity by varying a wide range of $\omega$ values. 
In the context of the $f(Q)$ model, at the WH throat where $r=r_0$, the energy density is valued at $160, \text{MeV/fm}^3$. This falls within the interval of $[41.58, 319.13], \text{MeV/fm}^3$, as $\omega$ varies widely within the range $\omega\in[0.3,1.0]$. In contrast, within the framework of GR, the corresponding value is on the order of $10^5, \text{MeV/fm}^3$, which significantly exceeds the relevant interval of $[41.58, 319.13]\, \text{MeV/fm}^3$, thereby showing the violation. In the third panel of Fig. \ref{f2}, we illustrate a 3D contour plot of $B(r)$ in $\text{MeV/fm}^3$, where both $r$ and $\omega(r)$ are varied. It is apparent that as $\omega$ increases, the value of $B$ at the throat decreases. However, $B$ tends towards zero as one moves away from the throat. Apart from that, like the previous case, the third panel of Fig. \ref{f2} demonstrates that as $\omega \rightarrow 0$, the bag pressure $B(r)$ increases, approaching its upper limit. This suggests that our constructed model predicts intensified quark interactions within the WH during the thermodynamic phase transition from the matter-dominated era to the radiation-dominated era. 
Next, we have determined the physical parameters of our constructed WH model for the above shape function, which are given below:
\begin{eqnarray}\label{4ba4}
   &&\hspace{0.0cm} \rho=\frac{\beta }{2}-\frac{a \alpha  b_1 r_0 }{r^3}\big(\frac{r}{r_0}\big)^{b_1}\\  \label{4ba5}
    &&\hspace{0cm} p_r=-\frac{\beta }{2}+\frac{1}{r^3 (r+r_0)}\Big[\alpha  r_0 \Big\{a (r-r_0) \big(\big(\frac{r}{r_0}\big)^{b_1}-1\big)\nonumber\\&&\hspace{1cm}+3 r-r_0\Big\}\Big]\\  \label{4ba6}
     &&\hspace{0cm}p_t= \frac{1}{2 r^3 (r+r_0)}\bigg[2 \alpha  r_0^2 \Big\{a \Big(\big(\frac{r}{r_0}\big)^{b_1}-1\Big)+1\Big\}-\beta  r^3 r_0\nonumber\\&&\hspace{1cm}+\alpha  r r_0 \Big\{a (b_1-1) \big(\frac{r}{r_0}\big)^{b_1}+a-3\Big\}-\beta  r^4\bigg]
\end{eqnarray}
Now, we shall examine the NEC mathematically using the above Eqs. (\ref{4ba4}-\ref{4ba6}) at WH throat. The mathematical expression of NEC is given by :
\begin{eqnarray}
    &&\hspace{0cm}\rho+p_r=\frac{1}{r^3 (r+r_0)}\big[\alpha  r_0 \big(-a ((b_1-1) r+b_1 r_0\nonumber\\&&\hspace{1.5cm}+r_0) (\frac{r}{r_0})^{b_1}-a r+a r_0+3 r-r_0\big)\big],\\
   &&\hspace{0cm} \rho+p_t=\frac{1}{2r^3 (r+r_0)}\Big[\alpha  r_0 \big(-a ((b_1+1) r+2 (b_1\nonumber\\&&\hspace{1.5cm} -1)r_0) \big(\frac{r}{r_0}\big)^{b_1}+a r-2 a r_0-3 r+2 r_0\big)\Big].
\end{eqnarray}
At the throat, i.e., $r=r_0$, the above equation reduces to
\begin{eqnarray}
   &&\hspace{0cm}\rho+p_r\Big|_{r=r_0} =\frac{\alpha -a \alpha  b_1}{r_0^2}\\
    &&\hspace{0cm}\rho+p_t\Big|_{r=r_0}=-\frac{3 a \alpha  b_1+\alpha }{4 r_0^2}
\end{eqnarray}
It is evident from the above expression that $\rho+p_r$ is a negative quantity. One can easily verify by substituting $\alpha=-0.0005$, $a=0.2$, $b_1=0.5$, and $r_0=1$ to the above equation. Further, one can check the violation of NEC in Fig. \ref{f3}. We have also plotted the energy density and SEC with the above choice of free parameters in Fig. \ref{f3}. It was noticed that energy density is positive in the entire spacetime, whereas SEC is violated in the neighborhood of the throat.

\subsubsection{Shape function: II}
Next, we have explored another model by utilizing a different well-established exponential shape function \cite{kd} :
\begin{eqnarray}\label{4bb1}
    b(r) = \frac{r}{\exp \big[\eta  \big(\frac{r}{r_0}-1\big)\big]}.
\end{eqnarray}
where $\eta$ is a constant parameters and $r_0$ is the radius of WH throat. 
Now for this shape function \eqref{4bb1}, the expression for $B(r)$ can be obtain from \eqref{4b1} as:
\begin{eqnarray}
    B(r)=\frac{e^{-(\eta  (\frac{r}{r_0}-1))}}{8 r^2 r_0 \omega  (r+r_0)}\Big[\beta  r^3 r_0 \omega  e^{\eta  (\frac{r}{r_0}-1)}+\beta  r^3 r_0 e^{\eta  (\frac{r}{r_0}-1)}\nonumber\\&&\hspace{-7.3cm}+2 \alpha  \eta  r^2 \omega +\beta  r^2 r_0^2 \omega  e^{\eta  (\frac{r}{r_0}-1)}+\beta  r^2 r_0^2 e^{\eta  (\frac{r}{r_0}-1)}\nonumber\\&&\hspace{-7.3cm}-4 \alpha  r_0^2 e^{\eta  (\frac{r}{r_0}-1)}+2 \alpha  \eta  r r_0 \omega -2 \alpha  r r_0 \omega -2 \alpha  r r_0\nonumber\\&&\hspace{-7.2cm}-2 \alpha  r_0^2 \omega +2 \alpha  r_0^2\Big]
\end{eqnarray}

\begin{figure*}

\includegraphics[width=5.5cm, height=4.1cm]{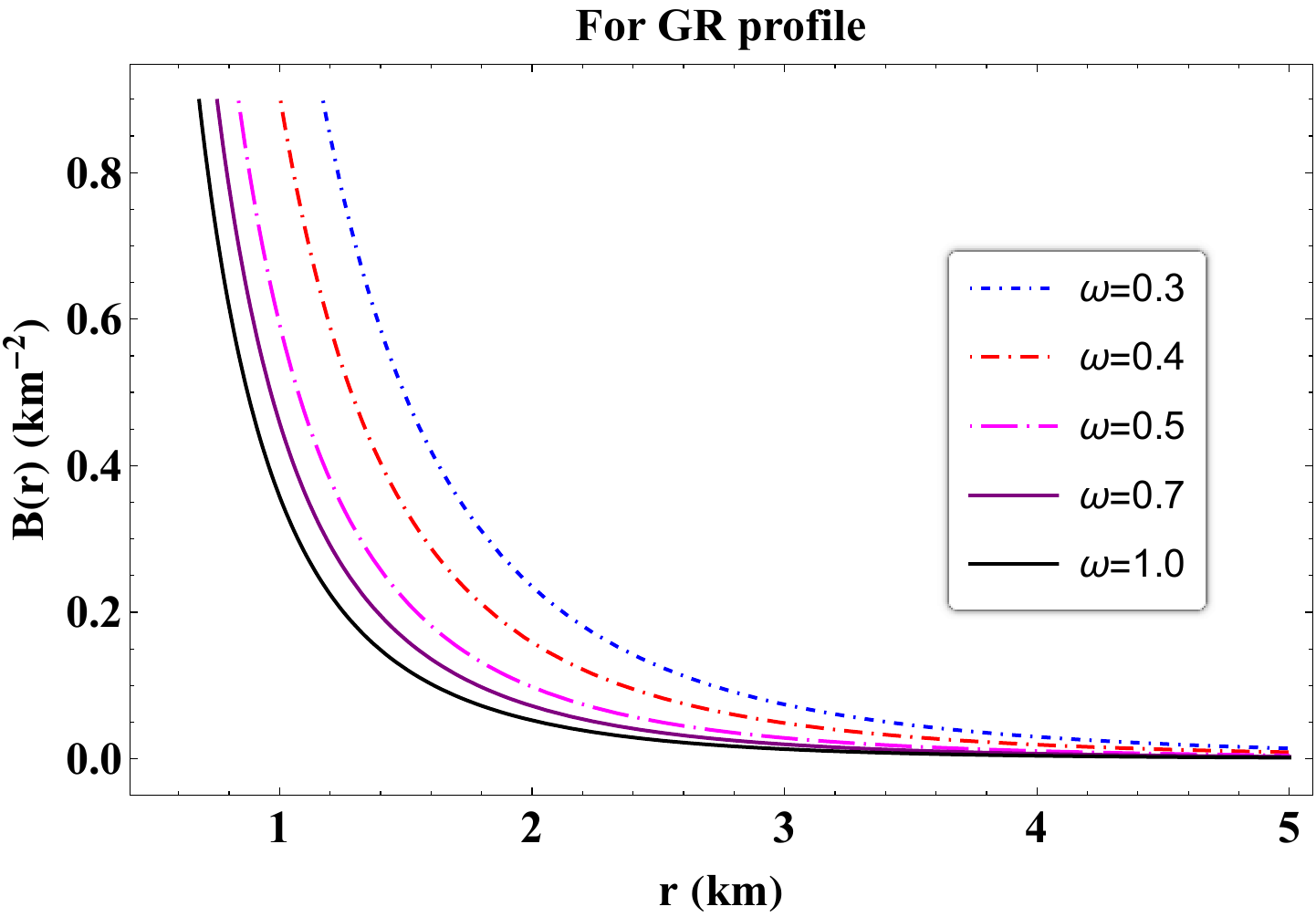}
\includegraphics[width=5.5cm, height=4.1cm]{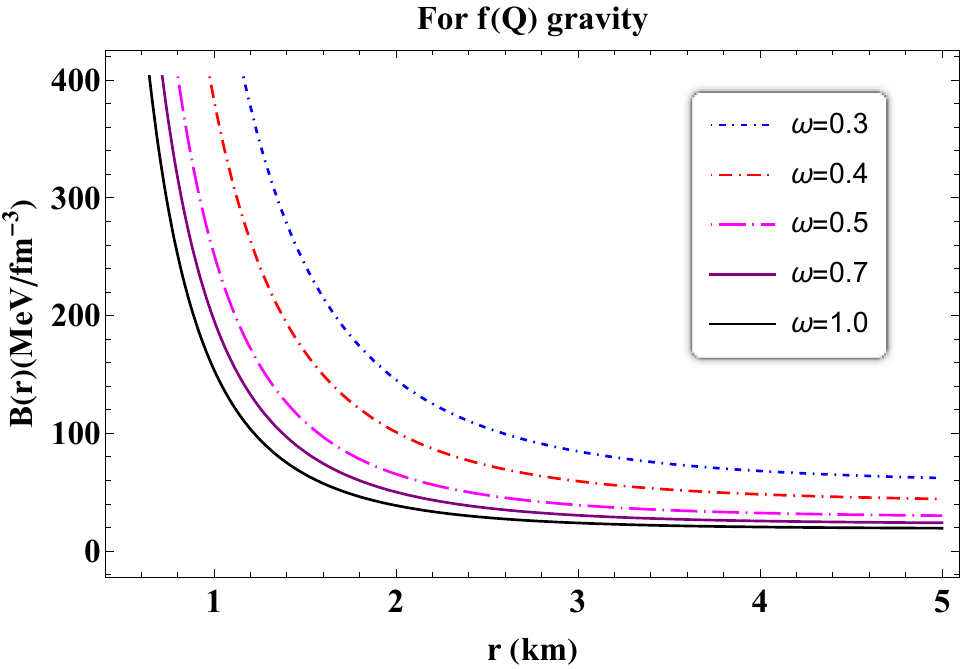} \includegraphics[width=6.0cm, height=4.2cm]{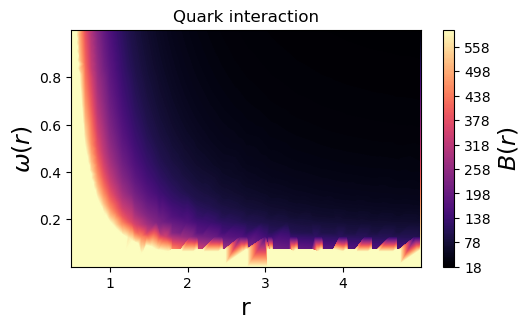}   
      \caption{(\textbf{Model-II, shape function II}): Left panel : Profile of $B(r)$ w.r.t. $'r'$ for GR ($\alpha=-1$, $\beta=0$, and $\eta=0.5$). Middle panel: Profile of $B(r)$ w.r.t. $r$ ($\alpha=-0.0005$, $\beta=0.0001$, and $\eta=0.5$). Right panel : Contour plot of $B(r)$ w.r.t. $r$ and $\omega$ ($\alpha=-0.0005$ $\beta=0.0001$, and $\eta=0.5$).} \label{f4}
\end{figure*}
\textbf{Physical analysis of $B(r)$ :} In Fig.\ref{f2}, we present a detailed graphical analysis of the density-dependent Bag parameter $B(r)$. In a similar way, we compared the WH model for GR and $f(Q)$ gravity by varying a broad range of $\omega$ values. Specifically, in the $f(Q)$ model at the WH throat ($r=r_0$), the value of $B$ is $400 \,\text{MeV/fm}^3$, but in this case, it slightly exceeds the interval $[41.58, 319.13]\,\text{MeV/fm}^3$ . Conversely, within the GR framework, the value of $B$ at the throat is around $0.8 \,\text{MeV/fm}^3$, significantly exceeding the interval $[41.58, 319.13]\,\text{MeV/fm}^3$, indicating an inconsequential. The third panel of Fig. \ref{f4} features a 3D contour plot of $B(r)$ in $\text{MeV/fm}^3$, with variations in both $r$ and $\omega(r)$. It is evident that as $\omega$ increases, the value of $B$ decreases. However, in the anisotropic WH model in this particular shape function (II), it gives a higher value of $B$ than the shape function (I). It is worth mentioning that from the three analysis of $B(r)$ it can be visible in the throat region of the WH it gives the higher value of $B$.  Moreover, as it approaches the higher radius, it is monotonically decreasing, indicating less interaction between the quark matter.

\begin{figure*}   
    \includegraphics[width=6.2cm, height=4.1cm]{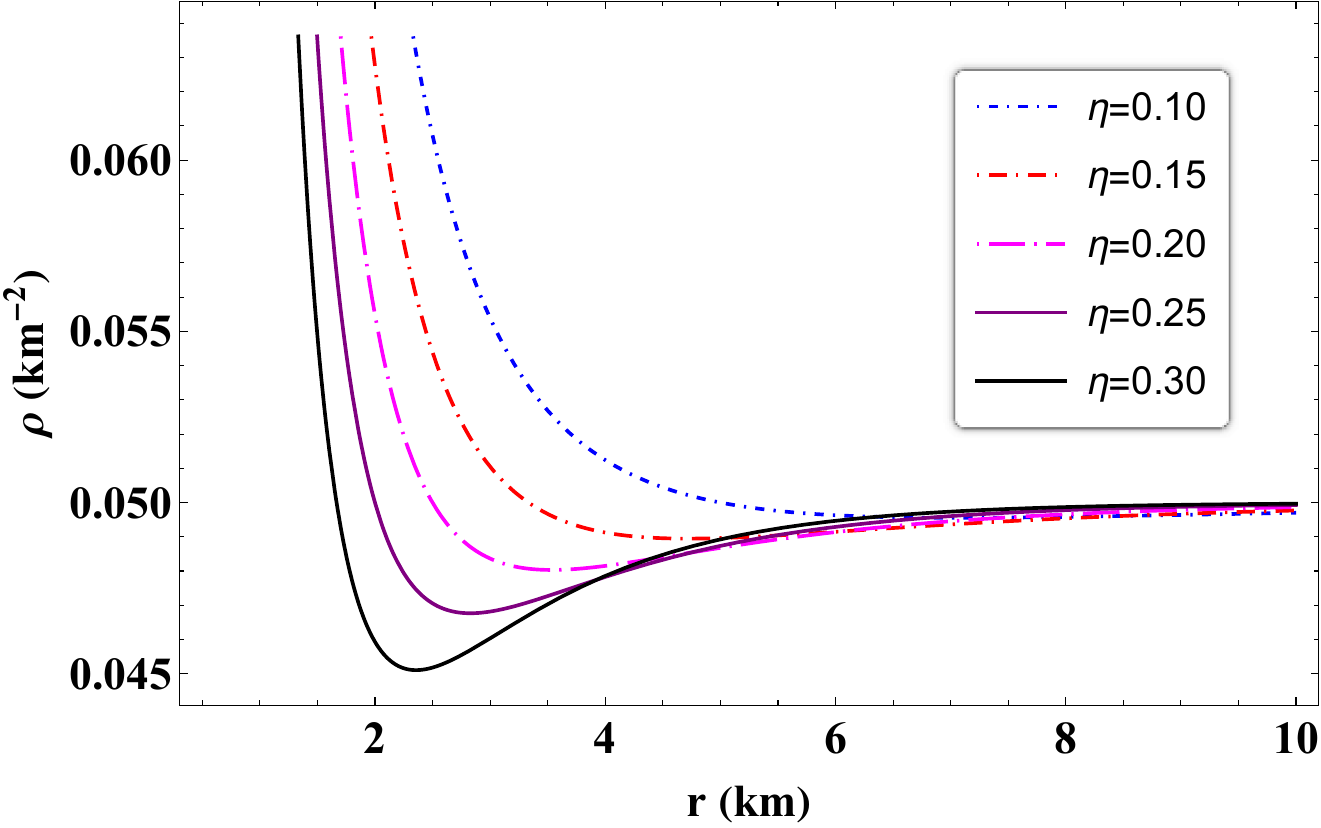}~
    \includegraphics[width=6.2cm, height=4.1cm]{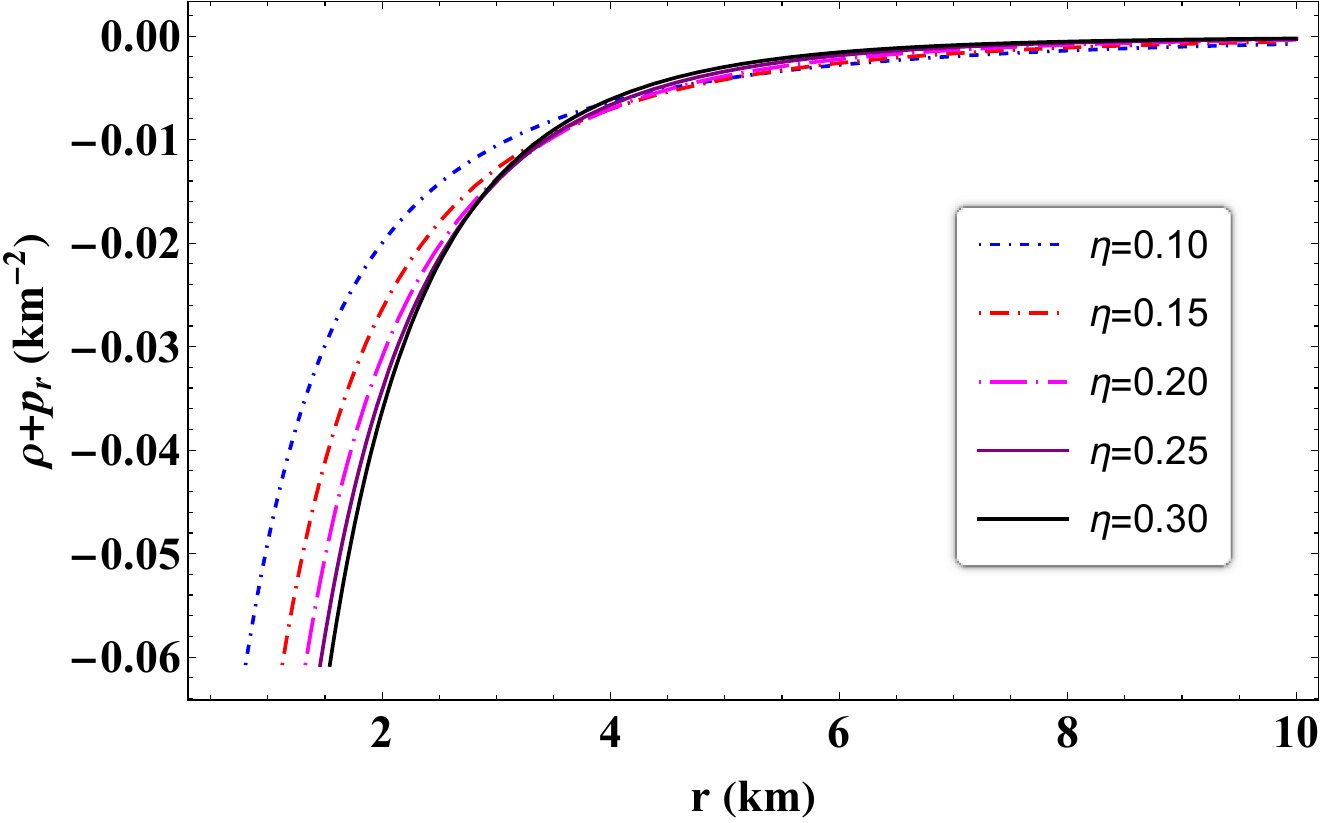}~
    \includegraphics[width=6.2cm, height=4.1cm]{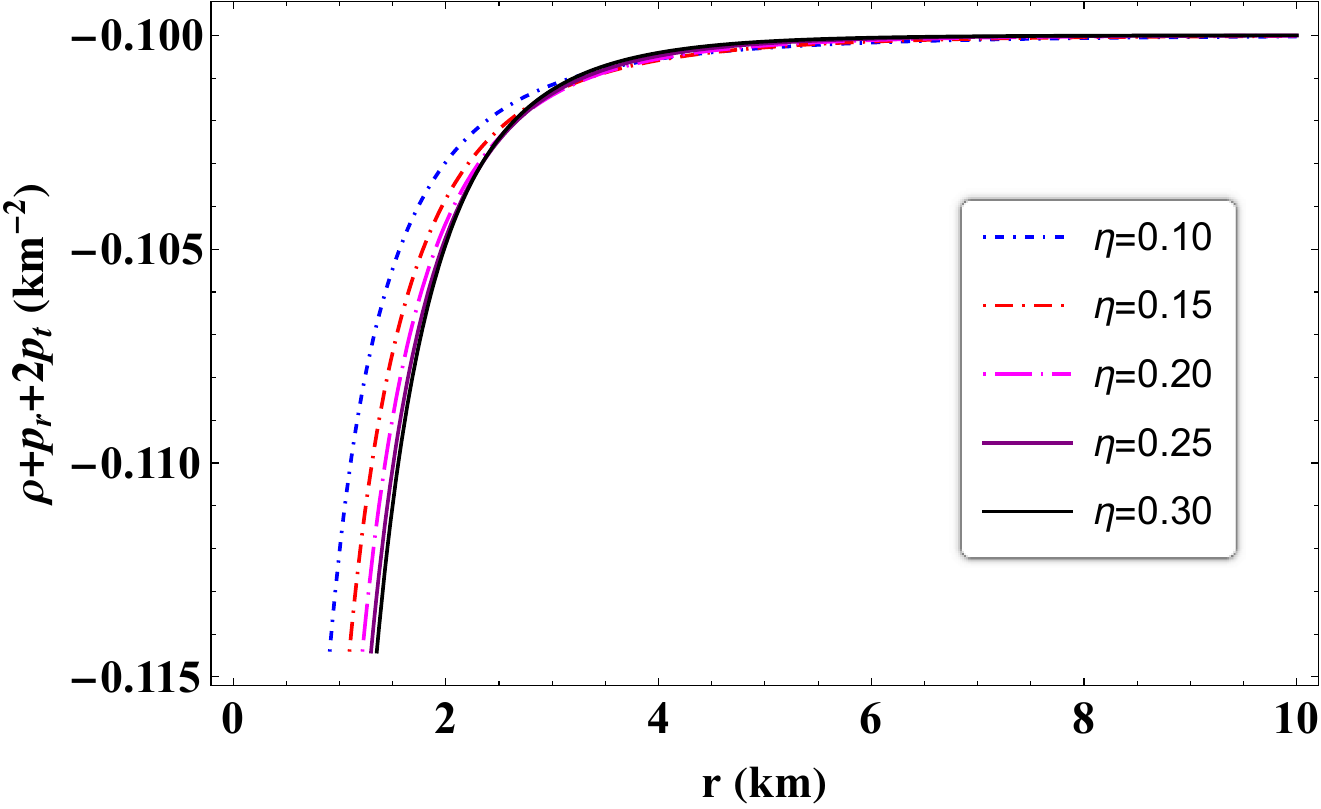}
       \caption{(\textbf{Model-II, shape function II}) Profile of various energy conditions w.r.t. $'r'$ for $\alpha=-0.0005$ $\beta=0.0001$, and $\eta=0.5$.}\label{f5}  
\end{figure*}

Now, we shall study the behavior of NEC under this particular shape function. For this case, the expressions for NEC can be obtained as
\begin{equation}
\rho+p_r= \frac{1}{\mathcal{K}_1}\Big[\alpha  \left(e^{\eta -\frac{\eta  r}{r_0}} \left(\eta  r (r+r_0)-2 r_0^2\right)+2 r_0^2\right)\Big],
\end{equation}
\begin{equation}
\rho+p_t= \frac{1}{2 \mathcal{K}_1}\Big[\alpha  \left(r e^{\eta -\frac{\eta  r}{r_0}} (\eta  r+2 (\eta -1) r_0)-2 r_0^2\right)\Big],
\end{equation}
where $\mathcal{K}_1=r^2 r_0 (r+r_0)$. Further, at $r=r_0$, the radial NEC, $\rho+p_r$ can be obtain as follow
\begin{equation}
\rho+p_r\Big|_{r=r_0} =  \frac{\alpha  \eta }{r_0^2},
\end{equation}
which is obviously negative for $\alpha<0$ and $\eta>0$. Now, by fixing $\alpha=-0.0005$, we checked the energy conditions for different values of $\eta$ in Fig. \ref{f5}. We have observed that energy density is positive, whereas NEC and SEC disrespect the energy conditions. It is interesting to note that as we increase the values of $\eta$, the contribution to the violation of $\rho+p_r$ becomes greater. This type of behavior may confirm the presence of exotic matter near the throat. For a thorough examination of the energy condition, one can refer to the table-\ref{table1}.\\

\begin{widetext}
 \begin{table*}[t]     
\caption{Behaviour of $\rho$, $\rho+p_r$, $\rho-|p_r|$, $\rho-|p_t|$, $\rho+p_r+2p_t$ w.r.t radial co-ordinate $r$, $\forall r\in[1,\infty]$ for quark matter supported WH}\label{table1}
\begin{tabular}{llllll}

\hline \hline
\multicolumn{6}{l}{~~~~~~~~~~~~~~~~~~~~~~~~~~~~~~~~~~~~~~~~~~~~~~~~~~~~~~~~~MIT WH Model-I: $b(r)=r_0+ar_0\big[\big(\frac{r}{r_0}\big)^{b_1}-1\big]$}    \\
\hline \hline
Energy conditions  & \multicolumn{5}{l}{~~~~~~~~~~~~~~~~~~~~~~~~~~~~~~~~~~~~~~~~~~~~~~~~~~~~~~Parameter value} \\ \hline
  &  $b_1=0.6$ &  $b_1=0.7$&$b_1=0.8$ & $b_1=0.9$ & $b_1=1.0$ \\ \hline
$\rho $& $\geq 0 ~ \forall ~r \in[r_0,\infty)$ & $\geq 0 ~ \forall ~r \in[r_0,\infty)$ & $\geq 0 ~ \forall ~r \in[r_0,\infty)$ & $\geq 0 ~ \forall ~r \in[r_0,\infty)$ & $\geq 0 ~ \forall ~r \in[r_0,\infty)$ \\

$\rho+p_r$&  $\leq 0 ~ \forall ~r \in[r_0,\infty)$ & $\leq 0 ~ \forall ~r \in[r_0,\infty)$ & $\leq 0 ~ \forall ~r \in[r_0,\infty)$ & $\leq 0 ~ \forall ~r \in[r_0,\infty)$ & $\leq 0 ~ \forall ~r \in[r_0,\infty)$  \\

 $\rho+p_t$  &  $\geq 0 ~ \forall ~r \in[r_0,\infty)$ & $\geq 0 ~ \forall ~r \in[r_0,\infty)$ & $\geq 0 ~ \forall ~r \in[r_0,\infty)$ & $\geq 0 ~ \forall ~r \in[r_0,\infty)$ & $\geq 0 ~ \forall ~r \in[r_0,\infty)$ \\
 
 $\rho+p_r+2p_t$ & $\leq 0 ~ \forall ~r \in[r_0,\infty)$ & $\leq 0 ~ \forall ~r \in[r_0,\infty)$ & $\leq 0 ~ \forall ~r \in[r_0,\infty)$ & $\leq 0 ~ \forall ~r \in[r_0,\infty)$ & $\leq 0 ~ \forall ~r \in[r_0,\infty)$   \\
 
  $\rho-|p_r|$& $\leq 0 ~ \forall ~r \in[r_0,\infty)$ & $\leq 0 ~ \forall ~r \in[r_0,\infty)$ & $\leq 0 ~ \forall ~r \in[r_0,\infty)$ & $\leq 0 ~ \forall ~r \in[r_0,\infty)$ & $\leq 0 ~ \forall ~r \in[r_0,\infty)$   \\
  
  $\rho-|p_t|$ &  $\geq 0 ~ \forall ~r \in[r_0,\infty)$ & $\geq 0 ~ \forall ~r \in[r_0,\infty)$ & $\geq 0 ~ \forall ~r \in[r_0,\infty)$ & $\geq 0 ~ \forall ~r \in[r_0,\infty)$ & $\geq 0 ~ \forall ~r \in[r_0,\infty)$ \\
\hline \hline
 \multicolumn{6}{l}{~~~~~~~~~~~~~~~~~~~~~~~~~~~~~~~~~~~~~~~~~~~~~~~~~~~~~~MIT WH Model-II: $b(r) = r\exp \big[1-\eta  \big(\frac{r}{r_0}\big)\big]$}   \\
\hline\hline
Energy conditions  & \multicolumn{5}{l}{~~~~~~~~~~~~~~~~~~~~~~~~~~~~~~~~~~~~~~~~~~~~~~~~~~~~~~Parameter value} \\ \hline
&  $\eta=0.10$ &  $\eta=0.15$&$\eta=0.20$ & $\eta=0.25$ & $\eta=0.30$ \\ \hline
   $\rho $&   $\geq 0 ~ \forall ~r \in[r_0,\infty)$ & $\geq 0 ~ \forall ~r \in[r_0,\infty)$ & $\geq 0 ~ \forall ~r \in[r_0,\infty)$ & $\geq 0 ~ \forall ~r \in[r_0,\infty)$ & $\geq 0 ~ \forall ~r \in[r_0,\infty)$  \\
  $\rho+p_r$&  $\leq 0 ~ \forall ~r \in[r_0,\infty)$ & $\leq 0 ~ \forall ~r \in[r_0,\infty)$ & $\leq 0 ~ \forall ~r \in[r_0,\infty)$ & $\leq 0 ~ \forall ~r \in[r_0,\infty)$ & $\leq 0 ~ \forall ~r \in[r_0,\infty)$  \\
 $\rho+p_t$  &  $\geq 0 ~ \forall ~r \in[r_0,\infty)$ & $\geq 0 ~ \forall ~r \in[r_0,\infty)$ & $\geq 0 ~ \forall ~r \in[r_0,\infty)$ & $\geq 0 ~ \forall ~r \in[r_0,\infty)$ & $\geq 0 ~ \forall ~r \in[r_0,\infty)$ \\
 $\rho+p_r+2p_t$ & $\leq 0 ~ \forall ~r \in[r_0,\infty)$ & $\leq 0 ~ \forall ~r \in[r_0,\infty)$ & $\leq 0 ~ \forall ~r \in[r_0,\infty)$ & $\leq 0 ~ \forall ~r \in[r_0,\infty)$ & $\leq 0 ~ \forall ~r \in[r_0,\infty)$   \\
  $\rho-|p_r|$& $\leq 0 ~ \forall ~r \in[r_0,\infty)$ & $\leq 0 ~ \forall ~r \in[r_0,\infty)$ & $\leq 0 ~ \forall ~r \in[r_0,\infty)$ & $\leq 0 ~ \forall ~r \in[r_0,\infty)$ & $\leq 0 ~ \forall ~r \in[r_0,\infty)$   \\
  $\rho-|p_t|$ &  $\geq 0 ~ \forall ~r \in[r_0,\infty)$ & $\geq 0 ~ \forall ~r \in[r_0,\infty)$ & $\geq 0 ~ \forall ~r \in[r_0,\infty)$ & $\geq 0 ~ \forall ~r \in[r_0,\infty)$ & $\geq 0 ~ \forall ~r \in[r_0,\infty)$ \\
\hline
\hline
\end{tabular}
\end{table*}
\end{widetext}
\section{WH solutions for phantom-like GCCG Model}\label{sec6}
The equation of state characterizing the generalized Chaplygin gas is expressed as \cite{Vg}:
\begin{eqnarray}
    p_{ch}=-\frac{A}{\rho_{ch}^{\alpha}},
\end{eqnarray}
where $A$ and $\alpha$ are positive constants, with $\alpha$ constrained within the range $0 < \alpha \leq 1$. The specific instance where $\alpha = 1$ corresponds to the Chaplygin gas. Notably, this model exhibits an intriguing behavior: during early cosmological epochs, its energy density mimics that of ordinary matter, scaling as $\rho_{ch} \approx a^{-3} $, whereas, in later stages, it resembles a cosmological constant, $\rho_{ch} = \text{const}$.\\
Here, we present some extensions to the cosmic Chaplygin gas model, which incorporate an adjustable initial parameter denoted as $\delta$. Specifically, we examine a generalized Chaplygin gas model whose equation of state converges to that of current Chaplygin unified models for dark matter and energy as $\delta$ approaches zero. Moreover, we ensure that this generalized model satisfies the following conditions: 

\begin{itemize}
    \item  It transitions to a de Sitter fluid in the late stages and when $\delta \to -1$.
    \item It converges to $p=\delta \rho$ in the limit where the Chaplygin parameter $A\to 0$.
    \item It reduces to Chaplygin gas model for considering $\delta\to 0,\gamma=1$.
    \item It also adopts the equation of state observed in current generalized Chaplygin unified dark matter models under the conditions $\delta\to 0$.
    
\end{itemize}
The GCCG is a theoretical model that combines dark matter and dark energy into a single fluid. This fluid generates negative pressure, which is crucial for the exotic matter needed in wormholes. The negative pressure balances the gravitational forces that would otherwise cause the wormhole to collapse. The GCCG helps keep the wormhole stable and open by providing this negative pressure, allowing objects to pass through.
An equation describing the state of matter, which has been demonstrated to fulfill all the aforementioned criteria from I to IV, includes \cite{m1,m2}:

\begin{eqnarray}\label{ch}
    p_r=-\rho ^{-\gamma } \Big[(\rho ^{\gamma +1}-\eta )^{-\delta }+\eta \Big]
\end{eqnarray}

Where $\eta=\frac{B}{1+\delta}-1$, $\gamma$, and $B$ are arbitrary constants that can take positive or negative values. The intriguing aspect of this group of Chaplygin gas models lies in their ability to embody phantom-like dark energy within a defined range of their constant parameters. In the subsequent analysis, we explore the existence of WH solutions and assess their stability under both constant and variable redshift functions.

\subsection{Zero tidal force WH}
In the initial approach, we investigate the WH paradigm employing a constant redshift function denoted as $\phi(r)=\text{constant}$, referred to as the zero tidal force WH model. For this constant redshift function the field equations (\ref{fe1}-\ref{fe3}) becomes
\begin{eqnarray}
 &&\hspace{-1.5cm} \rho =  \frac{\beta }{2}-\frac{\alpha  b'(r)}{r^2},\\
  &&\hspace{-1.5cm} p_r =\frac{\alpha  b(r)}{r^3}-\frac{\beta }{2},\\
 &&\hspace{-1.5cm} p_t= \frac{\alpha  (r b'(r)-b(r))}{2 r^3}-\frac{\beta }{2}.
\end{eqnarray}

\begin{figure}[h]
    \centering
   \includegraphics[width=6.5cm, height=4.1cm]{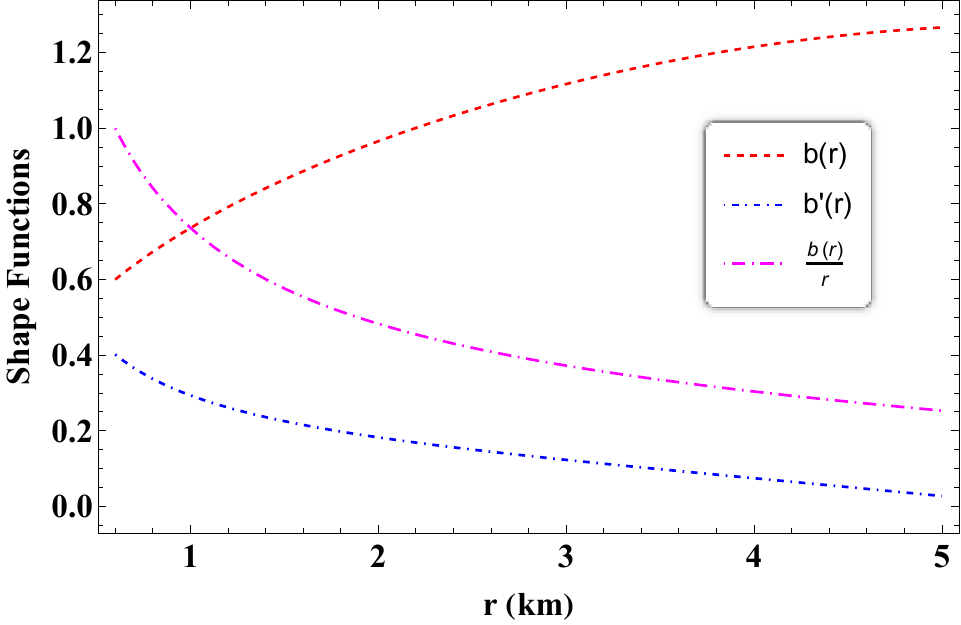}~
      \caption{ Shape function of tidal WH for GCCG model where we have taken the value of $\alpha=-0.1$, $\beta=0.001$,$\eta=-1$, $\delta =-1.6$  } \label{f222}  
\end{figure}

\begin{figure}
    \centering
    \includegraphics[scale=0.8]{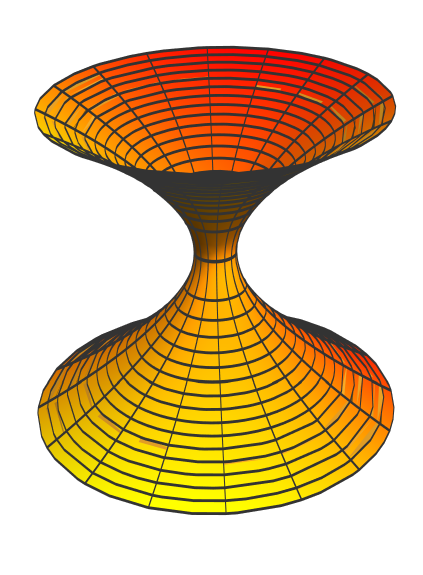}
    \caption{3D view of tidal WH.}
    \label{emb1}
\end{figure}

\begin{figure}[h]
    \centering
   \includegraphics[width=5.5cm, height=4.1cm]{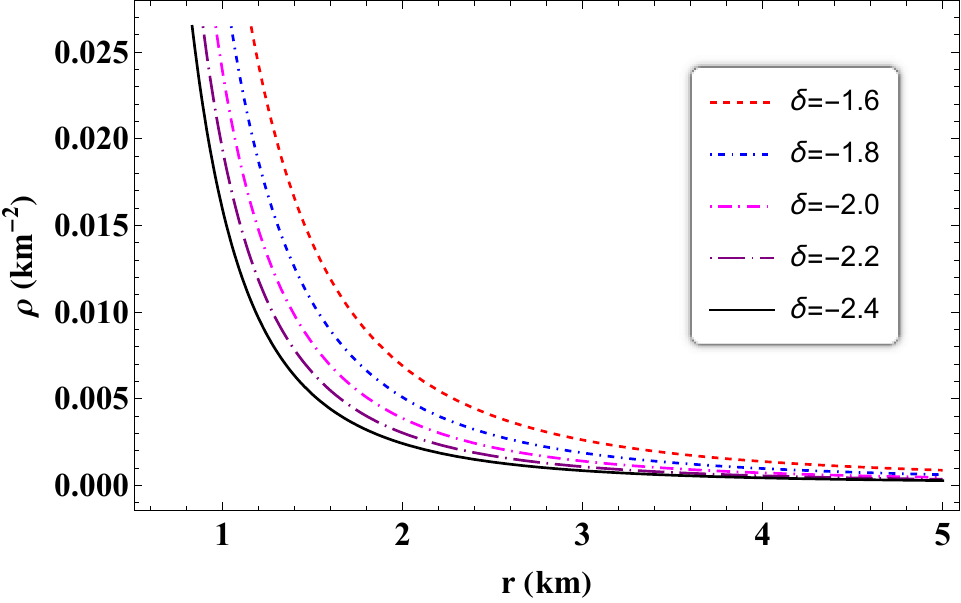}
    \includegraphics[width=5.5cm, height=4.1cm]{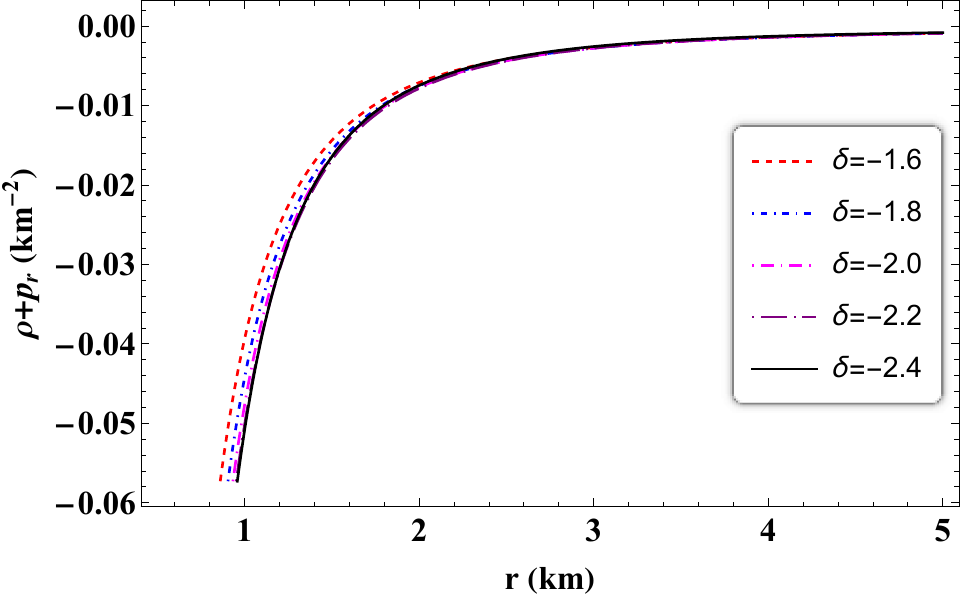}
      \caption{Energy condition profile of tidal WH for GCCG model. We have taken the value $\alpha=-0.1$, $\beta=0.001$,$\eta=-1$, } \label{f0000}  
\end{figure}

Now, with the help of these field equations and by utilizing the aforementioned equation of state (\ref{ch}), we get a first-order non-linear differential equation on shape function $b(r)$ as
\begin{eqnarray}
   \frac{\beta }{2}= \Big\{\frac{\beta }{2}-\frac{\alpha  b'(r)}{r^2}\Big\}^{-\gamma } \bigg[\Big\{\big(\frac{\beta }{2}-\frac{\alpha  b'(r)}{r^2}\big)^{\gamma +1}-\eta \Big\}^{-\delta }+\eta \bigg]\nonumber\\&&\hspace{-7.5cm}+\frac{\alpha  b(r)}{r^3}.
\end{eqnarray}
It is evident that the aforementioned differential equation poses a significant analytical challenge. Thus, to attain an analytical solution, we have opted for a specific constraint value, namely $\gamma=-1$. Subsequently, the expression for the shape function $b(r)$ is derived as presented below:
\begin{eqnarray}\label{5a5}
    b(r)=C_1 \left(  r \Gamma_2\right)^{\frac{(1-\eta )^{\delta }}{\Gamma_2}}+\frac{\beta  r^3  \left(\Gamma_2-(1-\eta )^{\delta }\right)}{2 \alpha \left(3 \Gamma_2-(1-\eta )^{\delta }\right) },~~~~
\end{eqnarray}
where $C_1$ is the integrating constant. Now, we impose the throat condition $b(r_0)=r_0$ to obtain $C_1$, and after substituting it to the above equation, we can get the final form of the shape function for zero tidal force WH
\begin{eqnarray}\label{5a7}
    b(r) = \frac{-1}{2 \alpha \mathcal{H}_2 }\bigg\{\beta  r^3 \left((1-\eta )^{\delta +1}-1\right)+r_0^{1-\Gamma_1} r^{\Gamma_1} \Big(-6 \alpha \nonumber\\&&\hspace{-8.3cm}+(1-\eta )^{\delta } \left(\alpha  (2-6 \eta )+\beta  (\eta -1) r_0^2\right)+\beta  r_0^2\Big)\bigg\}~~~~~
\end{eqnarray}
We have investigated the behavior of the above shape function graphically in Fig. \ref{f222}. It is clear that the shape function \eqref{5a7} respects the flare-out condition $b^{'}(r_0)<1$ under the asymptotic condition. 
Also, we have embedded the 3D diagram of the wormhole to visualize the wormhole, which can be found in \ref{emb1}.\\
Now, we can obtain the component of the energy-momentum tensor using the shape function \eqref{5a7}
\begin{eqnarray}
   \nonumber\\&&\hspace{0.4cm} \rho= \frac{1}{2 \mathcal{H}_2}\Big[(1-\eta )^{\delta } \big(2 \beta  +\frac{1}{r^3 \Gamma_2}\big\{r_0^{1-\Gamma_1} r^{\Gamma_1}  (-6 \alpha \nonumber\\&&\hspace{0.8cm} +(1-\eta )^{\delta } (\alpha  (2-6 \eta )+\beta  (\eta -1) r_0^2)+\beta  r_0^2)\big\}\big)\Big],~~~\\
    \nonumber\\&&\hspace{0.6cm} p_r=\frac{\left(\alpha  \Gamma_2 \right)^{-1}}{2r^3 \left((1-3 \eta ) (1-\eta )^{\delta }-3\right) }\bigg[\alpha  \Gamma_2\Big(\beta  r^3\nonumber\\&&\hspace{1.0cm} \left((1-\eta )^{\delta +1}-1\right)+r_0^{1-\Gamma_1}r^{\Gamma}  (-6 \alpha +(1-\eta )^{\delta }\nonumber\\&&\hspace{1.2cm} (\alpha  (2-6 \eta )+\beta  (\eta -1) r_0^2)+\beta  r_0^2)\Big)\bigg]-\frac{\beta}{2},\\
   \nonumber\\&&\hspace{0.7cm} p_t=\frac{1}{\mathcal{H}_1}\bigg[-4 \beta  \left(\Gamma_2\right)^2+r_0^{1-\Gamma_1}r^{-3+\Gamma_1} (1-\nonumber\\&&\hspace{1.5cm}(1-\eta )^{\delta +1}) \big(-6 \alpha +(1-\eta )^{\delta } \big(\alpha  (2-6 \eta )\nonumber\\&&\hspace{1.5cm}+\beta  (\eta -1) r_0^2\big)+\beta  r_0^2\big)\bigg],
 \end{eqnarray}
where\\
$\Gamma_1=^{\frac{1}{(1-\eta )^{-\delta }+\eta }}$,~~~$\Gamma_2=\eta  (1-\eta )^{\delta }+1$,\\
$\mathcal{H}_1=4 \left((6 \eta -1) (1-\eta )^{\delta }+\eta  (3 \eta -1) (1-\eta )^{2 \delta }+3\right)$,\\
$\mathcal{H}_2=\left((3 \eta-1 ) (1-\eta )^{\delta }-3\right)$.\\
Further, we checked NEC at $r=r_0$,
\begin{equation}
\rho+p_r\Big|_{r=r_0}=\frac{\left(1-(1-\eta )^{\delta +1}\right) \left(2 \alpha -\beta  r_0^2\right)}{2 r_0^2 \left(\Gamma_2\right)}<0
\end{equation}
We have presented the behavior of NEC in Fig. \ref{f0000}. In this case, we noticed that as we increased the negative values of $\delta$, the contribution to the violation of NEC became greater. We also studied energy density, which shows a positive decrease in behavior.
In the subsequent section, our focus lies on examining the WH model within the framework of a variable redshift function set against the backdrop of a cosmic Chaplygin gas exhibiting phantom-like characteristics.

\subsection{Models with variable redshift function}
In this section, we shall discuss WH solutions supported by GCCG with non-constant redshift functions.
\subsubsection{RF-I: $\phi(r)=-\frac{\chi}{r}$}
We first consider the redshift function of the form
\begin{eqnarray}\label{sh1}
\phi(r)=-\frac{\chi}{r},\, \chi>0
\end{eqnarray}
This redshift function under consideration demonstrates asymptotic flatness, indicating that, i.e., $\phi(r)\rightarrow 0$ as $r\rightarrow \infty$. Using the above redshift function by Kar and Sahdev \cite{Sahdev} delved into its implications on the behavior of matter (mainly focusing on the WEC) and how it affects the structure of spacetime slices. Subsequently, L. A. Anchordoqui and colleagues \cite{Anchordoqui} further examined WHs evolving with this redshift function, shedding light on the challenges associated with violating the WEC and the potential for human travel through these dynamic spacetime structures. Additionally, investigations into WH solutions using this redshift function have been conducted in various modified theories of gravity, including $f(R)$ gravity \cite{Fayyaz1}, $f(R,\Phi)$ gravity \cite{Fayyaz2}, and $f(Q)$ gravity \cite{Mustafa}.

\begin{figure}[h]
\centering
    \includegraphics[width=4.2cm, height=3.5cm]{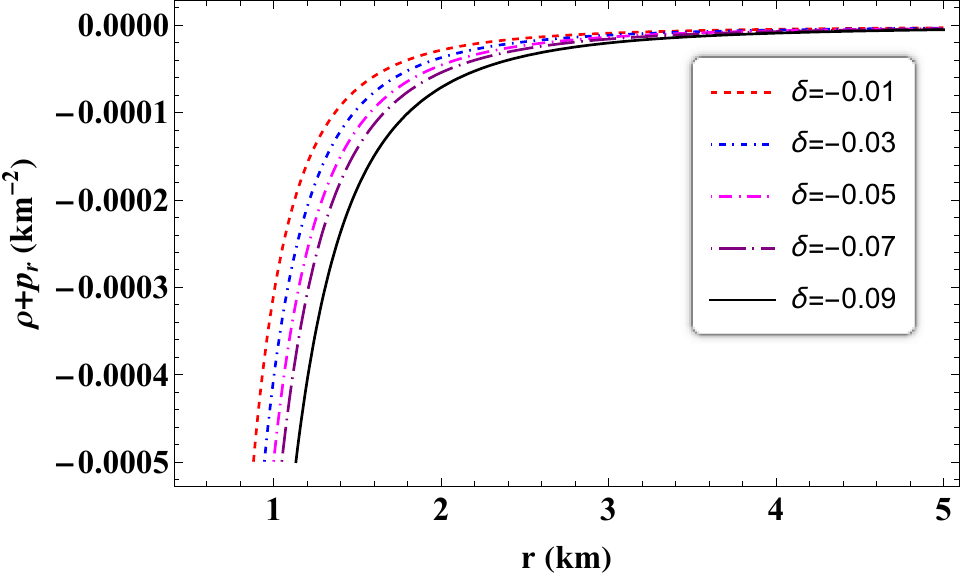}
    \includegraphics[width=4.2cm, height=3.5cm]{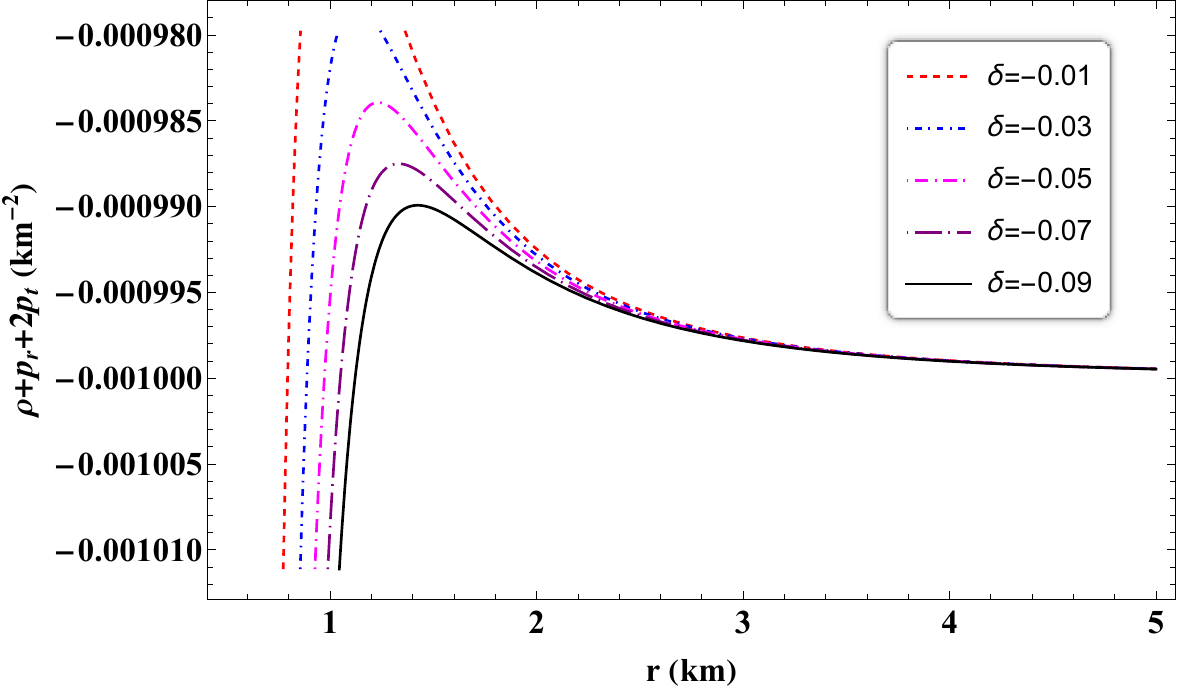}
\caption{Profile of energy conditions for GCCG WH model (RF-I) with $\alpha=-0.1$, $\beta=0.001$,$\eta=0.005$.}
\label{f11} 
\end{figure}

For the redshift function (\ref{sh1}), and by utilizing Eq. (\ref{fe1}) and GCCG (\ref{ch}), we get the shape function of the WH given below:
\begin{eqnarray}\label{a}
   &&\hspace{1cm} b(r)=\frac{1}{\alpha \Gamma_2^4}e^{\left(\frac{(1-\eta )^{\delta } (r \log (r)-2 \chi)}{r\Gamma_2}\right)}\Bigg\{\alpha  r_0 \Gamma_2^4 \exp [(r_0^{-1}(1-\eta )^{\delta} \nonumber\\&&\hspace{1cm} (2 \chi-r_0 \log (r_0))]-2 \alpha  \chi (1-\eta )^{\delta}\Gamma_2^3 r^{-\Gamma_1}\times\mathcal{I}_1\nonumber\\&&\hspace{1cm}+\frac{1}{2} \beta  \left(1-(1-\eta )^{\delta +1}\right) r^{3+\Gamma_1}\Gamma_2^3 \times \mathcal{I}_2+ 2^{1-\Gamma_1} \nonumber\\&&\hspace{1cm}(1-\eta )^{\delta } r_0^{-\Gamma_1} \left(-r_0^{-1}\chi (1-\eta )^{\delta }\right)^{-\Gamma_1}\big(\alpha \Gamma_2^3\mathcal{G}_1\nonumber\\&&\hspace{1cm}+2 \beta  \chi^2 \left(1-(1-\eta )^{\delta +1}\right) (1-\eta )^{2 \delta }\mathcal{G}_2\big)\Bigg\}.
\end{eqnarray}
 One can notice from the above-determined shape function that we have derived the shape function for a particular constrained value $\gamma=-1$; otherwise, for other $\gamma$ values, we can not get the analytical solution. Because of long-expression we have provided the value of $\mathcal{I}_1$, $\mathcal{I}_2$, $\mathcal{G}_1$ and $\mathcal{G}_2$ in the Appendix.\\
\begin{figure}[h]
    \centering
   \includegraphics[width=4.2cm, height=3.5cm]{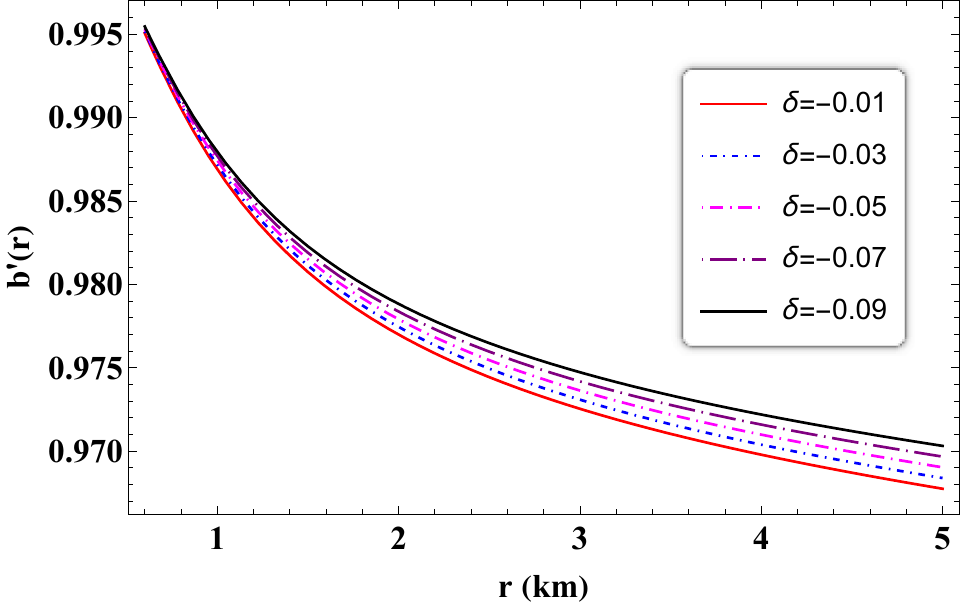}
    \includegraphics[width=4.2cm, height=3.5cm]{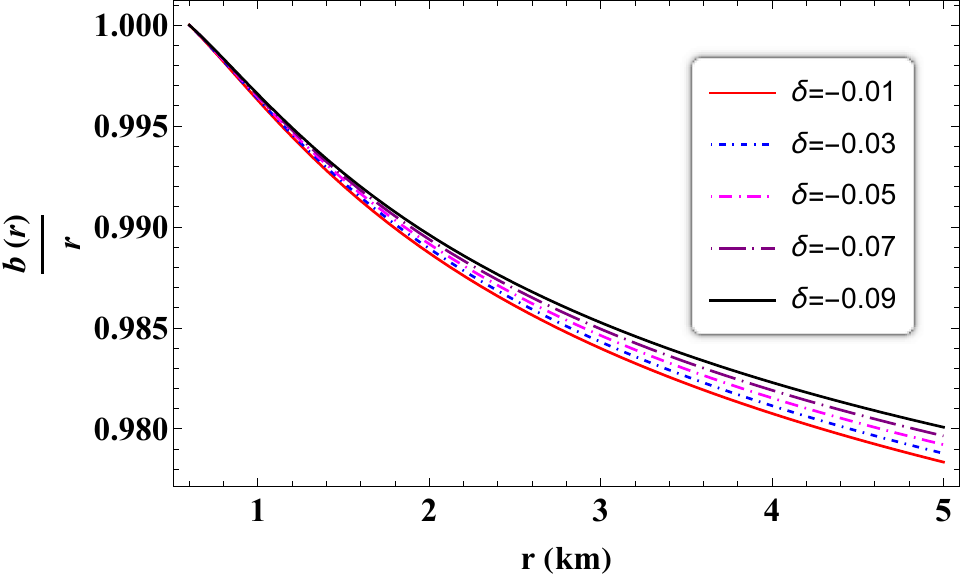}
\caption{Necessary condition on shape function for GCCG WH (RF-I) where $\alpha=-0.1$, $\beta=0.001$, $\eta=0.005$. }
\label{f00}  
\end{figure}
We have visually analyzed the properties of the shape functions, including the flare-out condition and asymptotic flatness condition, in Fig. \ref{f00}. To simplify the analysis, we chose the following parameters: $\alpha=-0.1$, $\beta=0.001$, $\eta=0.005$, $r_0=0.5$ and varied different $\delta$. 
From Fig. \ref{f00}, we observed that $\frac{b(r)}{r}\rightarrow 0$ as $r\rightarrow \infty$, confirming the asymptotic flatness condition. Additionally, the flare-out condition was satisfied at the WH throat. Now by implementing the redshift function (\ref{sh1}) and shape function (\ref{a}) one can get the components of energy-momentum tensor $\rho$, $p_r$ and $p_t$ from the Eq. (\ref{fe1}-\ref{fe3}). We have not given the full expression of those quantities for the long-expression.
We also examined the NEC and SEC in Fig. \ref{f11} using the same parameters discussed in the shape function. NEC is significant in energy conditions, as its violation could indicate the presence of exotic matter at the WH throat. Fig. \ref{f11} clearly shows a violation of the NEC, highlighting the impact of the parameters on NEC violation. Furthermore, we confirmed the violation of the SEC throughout the entire spacetime for different $\delta$ parameters.

\subsubsection{RF-II: $\phi(r)=\text{Log}\big(1+\frac{r_0}{r}\big)$}
We have formulated an alternative WH model supported by a phantom-like generalized Chaplygin gas, adopting a distinct redshift function \cite{phi1}
\begin{eqnarray}
    \phi(r)=\text{Log}\big(1+\frac{r_0}{r}\big)\label{sf2}
\end{eqnarray}
Note that the above redshift function satisfies the asymptotically flatness condition. Now, by using the Eqs. (\ref{fe1}) and (\ref{ch}), one can obtain the shape function $b(r)$ under the above redshift function, given by
\begin{eqnarray}\label{b}
  \nonumber\\&&\hspace{0cm}  b(r)=-\frac{1}{d_1}\exp \big(\frac{(1-\eta )^{\delta }  \log (\frac{(r+r_0)^2}{r})}{\Gamma_2}-\frac{(1-\eta )^{\delta } \log(4r_0)}{\Gamma_2}\big)\nonumber\\&&\hspace{1cm}\Big\{-r_0+\frac{1}{d_2}\big[e^{(1-\eta )^{\delta } \log(4r_0)}r_0^{1-\frac{(1-\eta )^{\delta }}{\eta  (1-\eta )^{\delta }+1}}\big[\beta  r_0^2 (\eta ^2 (1-\eta )^{2 \delta }\nonumber\\&&\hspace{1cm}+2 \eta  (1-\eta )^{\delta }-(1-\eta )^{2 \delta }+1)\times \mathcal{F}_1+4 \alpha  (1-\eta )^{\delta } \nonumber\\&&\hspace{1cm}\left(3 \eta  (1-\eta )^{\delta }+(1-\eta )^{\delta }+3\right)\times\mathcal{F}_2\big]\big]\Big\}
\end{eqnarray}
Because of long-expression we have provided the value of $d_1$, $d_2$, $\mathcal{F}_1$, $\mathcal{F}_2$, $\mathcal{F}_3$ and $\mathcal{F}_4$ in the Appendix.\\
It is noteworthy from the aforementioned derived shape function that we have obtained it under a specific constraint value $\gamma=1$. Conversely, for different values of $\gamma$, obtaining an analytical solution becomes unfeasible.\\

\begin{figure}[h]
    \centering
   \includegraphics[width=4.2cm, height=3.5cm]{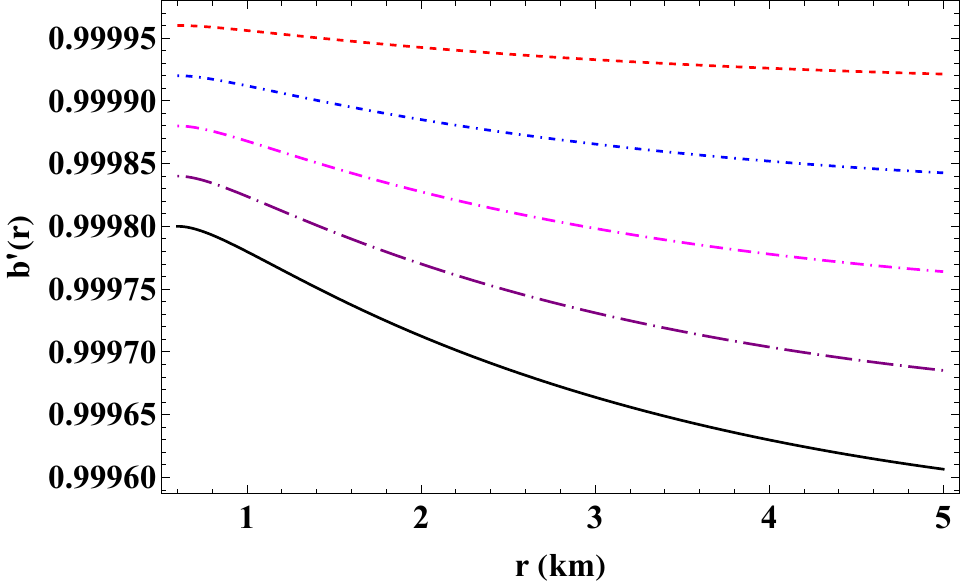}
    \includegraphics[width=4.2cm, height=3.5cm]{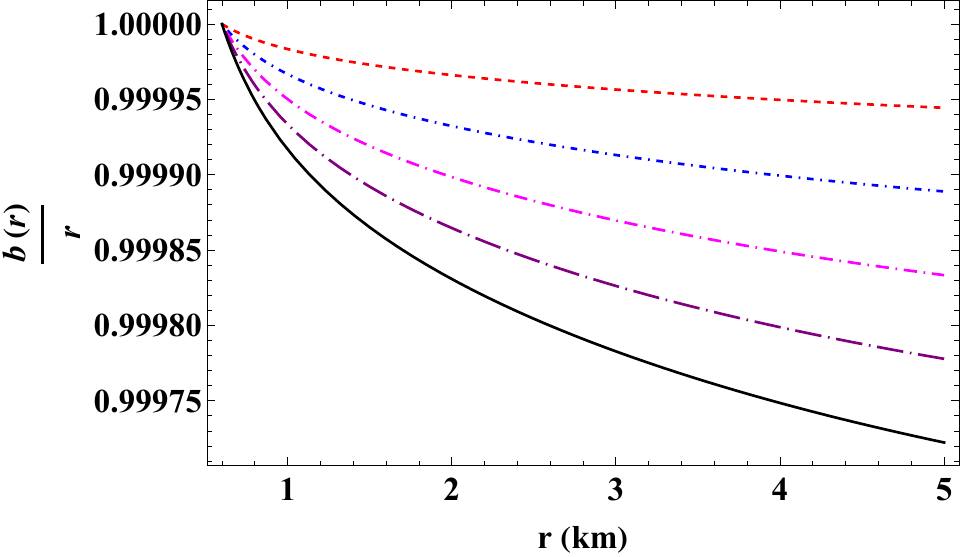}
      \caption{Necessary condition on shape function for GCCG (RF-II) WH where $\alpha=-0.1$, $\beta=0.001$, $\eta=0.005$ and $\delta=-1.01$, ($\textcolor{red}\star$),  $\delta=-1.02$,($\textcolor{blue}\star$), $\delta=-1.03$,($\textcolor{magenta}\star$), $\delta=-1.04$,($\textcolor{violet}\star$), and $\delta=-1.05$($\textcolor{black}\star$).} \label{f22}  
\end{figure}
\begin{figure}[h]
\centering
    \includegraphics[width=4.2cm, height=3.5cm]{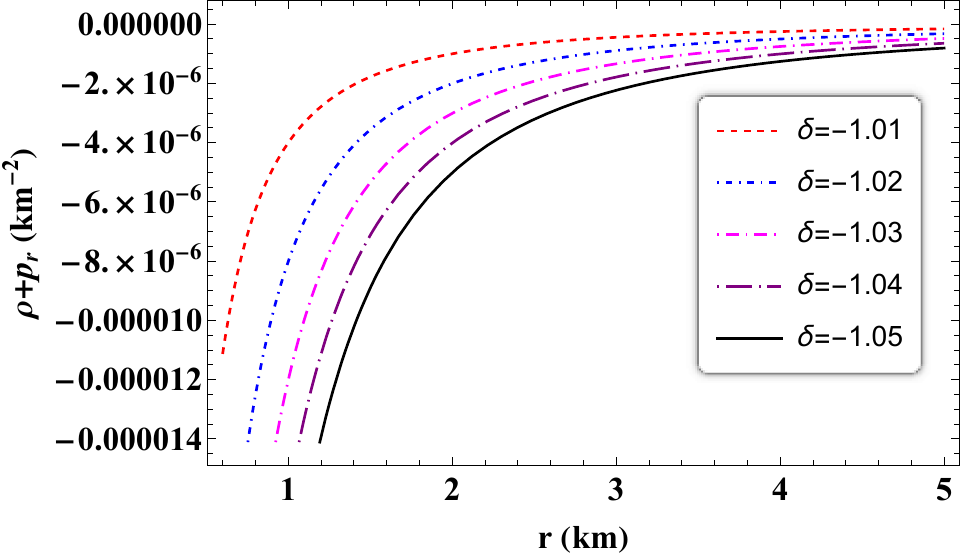}
    \includegraphics[width=4.2cm, height=3.5cm]{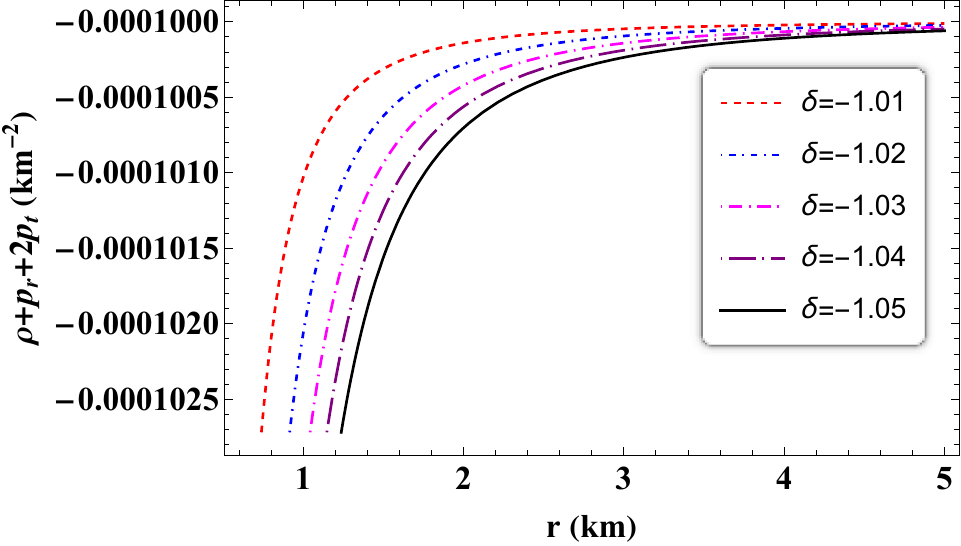}
\caption{Profile of energy conditions for GCCG WH model (RF-II) with $\alpha=-0.1$, $\beta=0.001$,$\eta=0.004$.} \label{f33}  
\end{figure}

In Fig. \ref{f22}, we have plotted $b'(r)$ and $\frac{b(r)}{r}$ with respect to $r$ using the free parameters $\alpha=-0.1, \beta=-0.001, \eta=-0.004$. We observed that at $r=r_0$, the condition $b'(r)<1$ confirms that the flare-out condition is satisfied at the throat. Additionally, we noticed that for high values of $\delta$, the flare-out condition may be violated in the WH throat. Furthermore, the asymptotically flat condition $\frac{b(r)}{r}\rightarrow 0$ is satisfied as $r\rightarrow \infty$. These favorable properties of the shape function are crucial for a WH to be traversable.  Now by implementing the redshift function (\ref{sf2}) and shape function (\ref{b}) one can get the physical quantity of the WH fluid like $\rho$, $p_r$ and $p_t$ from the Eq.(\ref{fe1}-\ref{fe3}). We have not given the full expression of those quantities for the long-expression. Next, we examined the behavior of energy conditions, particularly the NEC near the throat. It was observed that NEC is violated near the throat for negative values of the parameter $\delta$ (see Fig. \ref{f33}). However, for large values of $\delta$, the violation of NEC is no longer present. Also, we checked the SEC, as shown in Fig. \ref{f33}, and found that the SEC does not satisfy the energy conditions.\\
Further, a detailed summary of all the energy conditions under each redshift function has been analyzed and shown in Table-\ref{t2}.
\section{Stability Analysis}\label{sec7}
\begin{figure}[H]
\centering
    \includegraphics[width=4.2cm, height=3.5cm]{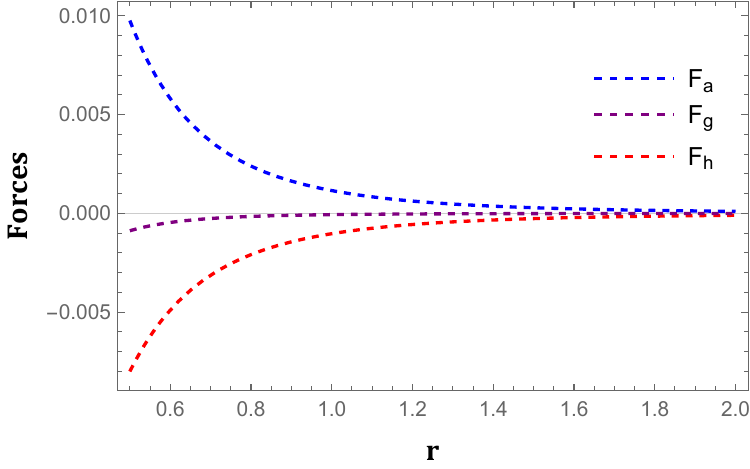}
    \includegraphics[width=4.2cm, height=3.5cm]{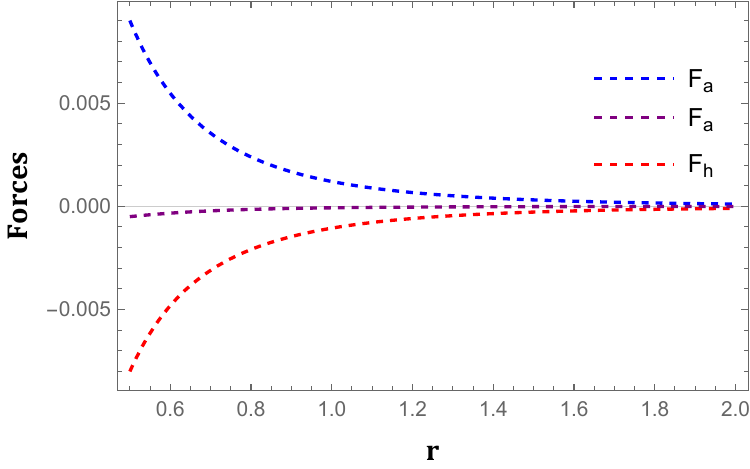}
\caption{Stability analysis for MIT bag model WH models  with $\alpha=-0.1$, $\beta=0.001$,$\eta=0.004$.} \label{f34}  
\end{figure}

\begin{figure*}
    \includegraphics[width=5.2cm, height=4.1cm]{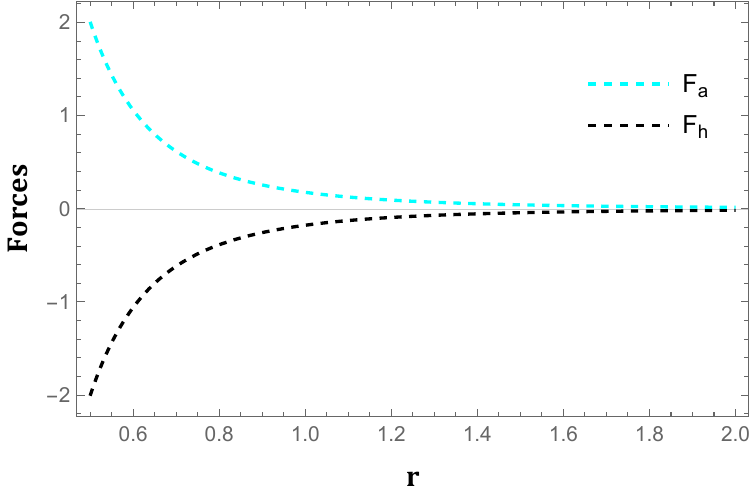}~
    \includegraphics[width=5.2cm, height=4.1cm]{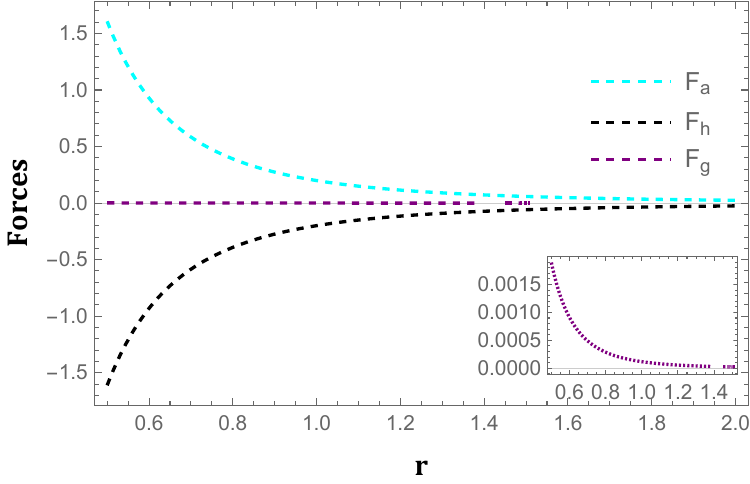}~
    \includegraphics[width=5.2cm, height=4.1cm]{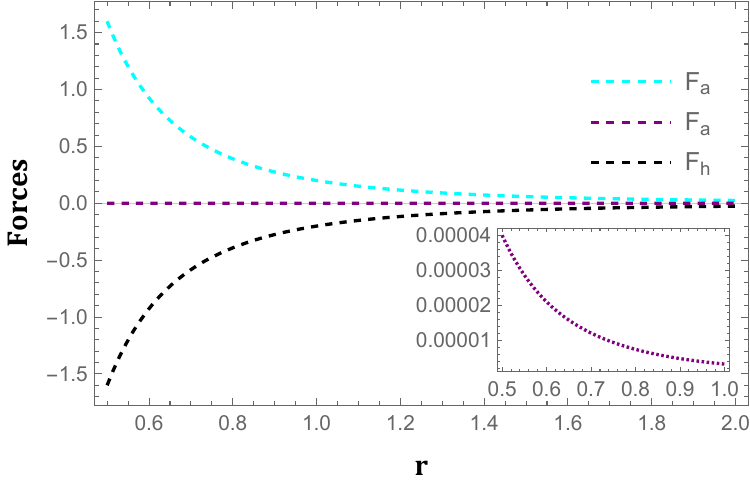}
       \caption{Stability analysis of GCCG WH w.r.t. $'r'$ Left: Tidal WH model, Middle: Variable RS-I model, Right: Variable RS-II model.
       }\label{f35}  
\end{figure*}

In this particular section, we consider the generalized Tolman-Oppenheimer-Volkoff (TOV) equation to investigate the stability of our obtained WH solutions. One straightforward method to derive the TOV equation involves using energy conservation principles, specifically by stating that the divergence of the stress-energy tensor must be zero, i.e., $\nabla_{\nu}T^{\mu\nu}=0$. The generalized TOV equation can be define as \cite{Islam,Kuhfittig/2020}
\begin{equation}\label{5a}
-\frac{dP_{r}}{dr}-\frac{\phi^{'}(r)}{2}(\rho+P_{r})+\frac{2}{r}(P_{t}-P_{r})=0,
\end{equation}
Naturally, the generalized TOV equation \eqref{5a} gives the information of the equilibrium condition for the WH
subject to the hydrostatic $(F_H)$ and gravitational $(F_G)$ force with another force called anisotropic $(F_A)$ force (because of anisotropic matter).\\
Hence, the equation \eqref{5a} takes the following form
\begin{equation}\label{5a2}
    F_H+F_G+F_A=0,
\end{equation}
where,
\begin{equation}\label{5a3}
F_h=-\frac{dP_{r}}{dr},    
\end{equation}
\begin{equation}\label{5a4}
F_g=-\frac{\phi^{'}}{2}(\rho+P_{r}), 
\end{equation}
\begin{equation}\label{5a5}
 F_a=\frac{2}{r}(P_{t}-P_{r}).   
\end{equation}
Based on the appropriate expressions, we plotted the graphs for the hydrostatic ($F_h$), gravitational ($F_g$), and anisotropic ($F_a$) forces for MIT bag model WH solutions in Fig. \ref{f34} and for GCCG WHs in Fig. \ref{f35}. Our observations revealed that in the MIT bag model case, the anisotropic force ($F_a$) exhibits positive behavior, whereas the hydrostatic force ($F_h$) and gravitational force ($F_g$) show negative behaviors. On the other hand, for the GCCG case, both $F_a$ and $F_g$ show positive behavior, while $F_h$ shows the opposite. Additionally, the combined effect of these forces provides insight into the system's stability and whether it is in equilibrium. This suggests that our constructed WH model is physically viable. For further details on this topic, refer to the following Refs. \cite{Karar,Ilyas}.

\begin{widetext}
 \begin{table*}
\caption{Behaviour of $\rho$, $\rho+p_r$, $\rho-|p_r|$, $\rho-|p_t|$, $\rho+p_r+2p_t$ w.r.t radial co-ordinate $r$, $\forall r\in[r_0,\infty]$ for GCCG WH}\label{t2}
\begin{tabular}{llllll}

\hline \hline
\multicolumn{6}{l}{~~~~~~~~~~~~~~~~~~~~~~~~~~~~~~~~~~~~~~~~~~~~~~~~~~~~~~~~~~~~~~~~~~~~~~~GCCG WH Model-I~~: ~~~~$\phi(r)=1$}    \\
\hline \hline
Energy conditions  & \multicolumn{5}{l}{~~~~~~~~~~~~~~~~~~~~~~~~~~~~~~~~~~~~~~~~~~~~~~~~~~~~~~Parameter value} \\ \hline
  &  $\delta=-1.6$ &  $\delta=-1.8$&$\delta=-2.0$ & $\delta=-2.2$ & $\delta=-2.4$ \\ \hline
$\rho $& $\geq 0 ~ \forall ~r \in[r_0,\infty)$ & $\geq 0 ~ \forall ~r \in[r_0,\infty)$ & $\geq 0 ~ \forall ~r \in[r_0,\infty)$ & $\geq 0 ~ \forall ~r \in[r_0,\infty)$ & $\geq 0 ~ \forall ~r \in[r_0,\infty)$ \\

$\rho+p_r$&  $\leq 0 ~ \forall ~r \in[r_0,\infty)$ & $\leq 0 ~ \forall ~r \in[r_0,\infty)$ & $\leq 0 ~ \forall ~r \in[r_0,\infty)$ & $\leq 0 ~ \forall ~r \in[r_0,\infty)$ & $\leq 0 ~ \forall ~r \in[r_0,\infty)$  \\

 $\rho+p_t$  &  $\geq 0 ~ \forall ~r \in[r_0,\infty)$ & $\geq 0 ~ \forall ~r \in[r_0,\infty)$ & $\geq 0 ~ \forall ~r \in[r_0,\infty)$ & $\geq 0 ~ \forall ~r \in[r_0,\infty)$ & $\geq 0 ~ \forall ~r \in[r_0,\infty)$ \\
 
 $\rho+p_r+2p_t$ & $\leq 0 ~ \forall ~r \in[r_0,\infty)$ & $\leq 0 ~ \forall ~r \in[r_0,\infty)$ & $\leq 0 ~ \forall ~r \in[r_0,\infty)$ & $\leq 0 ~ \forall ~r \in[r_0,\infty)$ & $\leq 0 ~ \forall ~r \in[r_0,\infty)$   \\
 
  $\rho-|p_r|$& $\leq 0 ~ \forall ~r \in[r_0,\infty)$ & $\leq 0 ~ \forall ~r \in[r_0,\infty)$ & $\leq 0 ~ \forall ~r \in[r_0,\infty)$ & $\leq 0 ~ \forall ~r \in[r_0,\infty)$ & $\leq 0 ~ \forall ~r \in[r_0,\infty)$   \\
  
  $\rho-|p_t|$ &  $\geq 0 ~ \forall ~r \in[r_0,\infty)$ & $\geq 0 ~ \forall ~r \in[r_0,\infty)$ & $\geq 0 ~ \forall ~r \in[r_0,\infty)$ & ${\leq 0 ~ \forall ~r \in[r_0,2.5]}$ & ${\leq 0 ~ \forall ~r \in[r_0,3]}$ \\ & & & & and ${\geq 0  ~\forall~ \in[2.5,\infty]}$ & and${\geq 0 ~ \forall ~r \in[3,\infty]}$ \\
\hline \hline
 \multicolumn{6}{l}{~~~~~~~~~~~~~~~~~~~~~~~~~~~~~~~~~~~~~~~~~~~~~~~~~~~~~~~~~~~~~~~GCCG WH Model-II~~: ~~~~~~$\phi(r)=-\frac{\chi}{r}$}   \\
\hline\hline
Energy conditions  & \multicolumn{5}{l}{~~~~~~~~~~~~~~~~~~~~~~~~~~~~~~~~~~~~~~~~~~~~~~~~~~~~~~Parameter value}\\ \hline
  &  $\delta=-0.01$ &  $\delta=-0.02$&$\delta=-0.03$ & $\delta=-0.04$ & $\delta=-0.05$ \\ \hline
   $\rho $&   $\geq 0 ~ \forall ~r \in[r_0,\infty)$ & $\geq 0 ~ \forall ~r \in[r_0,\infty)$ & $\geq 0 ~ \forall ~r \in[r_0,\infty)$ & $\geq 0 ~ \forall ~r \in[r_0,\infty)$ & $\geq 0 ~ \forall ~r \in[r_0,\infty)$  \\
  $\rho+p_r$&  $\leq 0 ~ \forall ~r \in[r_0,\infty)$ & $\leq 0 ~ \forall ~r \in[r_0,\infty)$ & $\leq 0 ~ \forall ~r \in[r_0,\infty)$ & $\leq 0 ~ \forall ~r \in[r_0,\infty)$ & $\leq 0 ~ \forall ~r \in[r_0,\infty)$  \\
 $\rho+p_t$  &  $\geq 0 ~ \forall ~r \in[r_0,\infty)$ & $\geq 0 ~ \forall ~r \in[r_0,\infty)$ & $\geq 0 ~ \forall ~r \in[r_0,\infty)$ & $\geq 0 ~ \forall ~r \in[r_0,\infty)$ & $\geq 0 ~ \forall ~r \in[r_0,\infty)$ \\
 $\rho+p_r+2p_t$ & $\leq 0 ~ \forall ~r \in[r_0,\infty)$ & $\leq 0 ~ \forall ~r \in[r_0,\infty)$ & $\leq 0 ~ \forall ~r \in[r_0,\infty)$ & $\leq 0 ~ \forall ~r \in[r_0,\infty)$ & $\leq 0 ~ \forall ~r \in[r_0,\infty)$   \\
  $\rho-|p_r|$& $\leq 0 ~ \forall ~r \in[r_0,\infty)$ & $\leq 0 ~ \forall ~r \in[r_0,\infty)$ & $\leq 0 ~ \forall ~r \in[r_0,\infty)$ & $\leq 0 ~ \forall ~r \in[r_0,\infty)$ & $\leq 0 ~ \forall ~r \in[r_0,\infty)$   \\
  $\rho-|p_t|$ &  $\geq 0 ~ \forall ~r \in[r_0,\infty)$ & $\geq 0 ~ \forall ~r \in[r_0,\infty)$ & $\geq 0 ~ \forall ~r \in[r_0,\infty)$ & $\geq 0 ~ \forall ~r \in[r_0,\infty)$ & $\geq 0 ~ \forall ~r \in[r_0,\infty)$ \\
\hline
\hline

 \multicolumn{6}{l}{~~~~~~~~~~~~~~~~~~~~~~~~~~~~~~~~~~~~~~~~~~~~~~~~~~~~~~~~~~~~~~~GCCG WH Model-III~~: ~~~~~~$\phi(r)=Log(1+\frac{r_0}{r})$}   \\
\hline\hline
Energy conditions  & \multicolumn{5}{l}{~~~~~~~~~~~~~~~~~~~~~~~~~~~~~~~~~~~~~~~~~~~~~~~~~~~~~~Parameter value}\\ \hline
  &  $\delta=-1.01$ &  $\delta=-1.02$&$\delta=-1.03$ & $\delta=-1.04$ & $\delta=-1.05$ \\ \hline
   $\rho $&   $\geq 0 ~ \forall ~r \in[r_0,\infty)$ & $\geq 0 ~ \forall ~r \in[r_0,\infty)$ & $\geq 0 ~ \forall ~r \in[r_0,\infty)$ & $\geq 0 ~ \forall ~r \in[r_0,\infty)$ & $\geq 0 ~ \forall ~r \in[r_0,\infty)$  \\
  $\rho+p_r$&  $\leq 0 ~ \forall ~r \in[r_0,\infty)$ & $\leq 0 ~ \forall ~r \in[r_0,\infty)$ & $\leq 0 ~ \forall ~r \in[r_0,\infty)$ & $\leq 0 ~ \forall ~r \in[r_0,\infty)$ & $\leq 0 ~ \forall ~r \in[r_0,\infty)$  \\
 $\rho+p_t$  &  $\geq 0 ~ \forall ~r \in[r_0,\infty)$ & $\geq 0 ~ \forall ~r \in[r_0,\infty)$ & $\geq 0 ~ \forall ~r \in[r_0,\infty)$ & $\geq 0 ~ \forall ~r \in[r_0,\infty)$ & $\geq 0 ~ \forall ~r \in[r_0,\infty)$ \\
 $\rho+p_r+2p_t$ & $\leq 0 ~ \forall ~r \in[r_0,\infty)$ & $\leq 0 ~ \forall ~r \in[r_0,\infty)$ & $\leq 0 ~ \forall ~r \in[r_0,\infty)$ & $\leq 0 ~ \forall ~r \in[r_0,\infty)$ & $\leq 0 ~ \forall ~r \in[r_0,\infty)$   \\
  $\rho-|p_r|$& $\leq 0 ~ \forall ~r \in[r_0,\infty)$ & $\leq 0 ~ \forall ~r \in[r_0,\infty)$ & $\leq 0 ~ \forall ~r \in[r_0,\infty)$ & $\leq 0 ~ \forall ~r \in[r_0,\infty)$ & $\leq 0 ~ \forall ~r \in[r_0,\infty)$   \\
  $\rho-|p_t|$ &  $\geq 0 ~ \forall ~r \in[r_0,\infty)$ & $\geq 0 ~ \forall ~r \in[r_0,\infty)$ & $\geq 0 ~ \forall ~r \in[r_0,\infty)$ & $\geq 0 ~ \forall ~r \in[r_0,\infty)$ & $\geq 0 ~ \forall ~r \in[r_0,\infty)$ \\
\hline
\hline
\end{tabular}
\end{table*}
\end{widetext}

\section{Volume Integral Quantifier}\label{sec8}
The ``total amount of exotic matter" that is necessary for the wormhole to be maintained is specified by the volume integral quantifier measurement. Within the spherically symmetric spacetime, the volume integral quantifier can be measured by the formula
\begin{equation}
    I_{v}=\int_{l_0}^{\infty} \int_{0}^{\pi} \int_{0}^{2 \pi} (\rho+p_r) \sqrt{-g} \,dr \,d\theta \, d\phi.
\end{equation}
After some laborious calculation, one might arrive at the curvilinear integral solution to the aforementioned equation by simplifying it, which is exactly represented by the formula below

\begin{equation}
    I_{v}= \oint [\rho+p_r] dV,
\end{equation}
where $dV=r^2 sin\theta \,dr \, d\theta \, d\phi.$
It can be expressed as
\begin{equation}
    I_{v}=8 \pi \int_{r_0}^{\infty} [\rho+p_r] \, r^2 dr.
\end{equation}
Let us imagine the wormhole extends from the throat, $r_0$, with the stress-energy tensor cutting off at a certain radius $r_1$, at which point it simplifies to
\begin{equation}
    I_{v}=8 \pi \int_{r_0}^{r_1} [\rho+p_r] \, r^2 dr.
\end{equation}
\begin{figure}[h]
    \centering
    \includegraphics[scale=0.6]{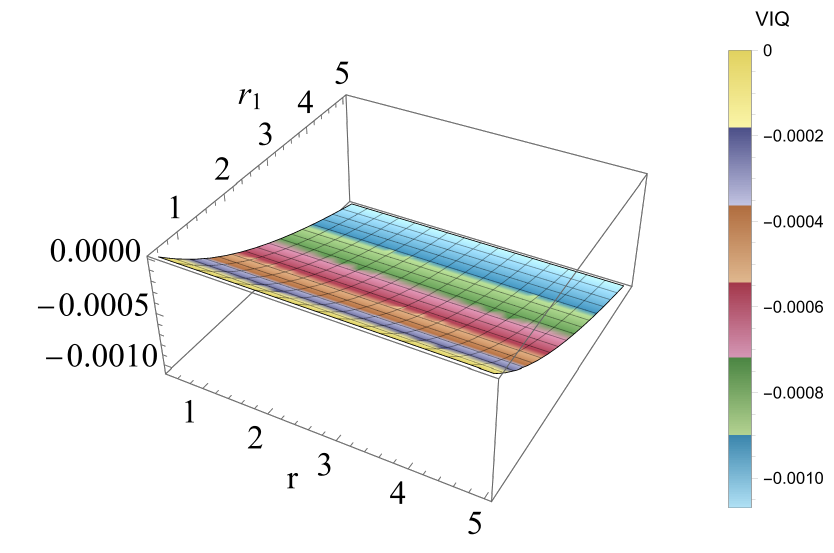}
    \includegraphics[scale=0.6]{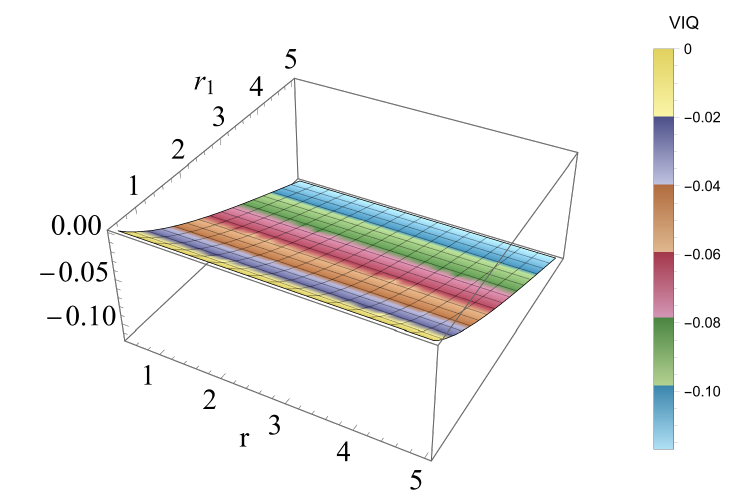}
    \includegraphics[scale=0.6]{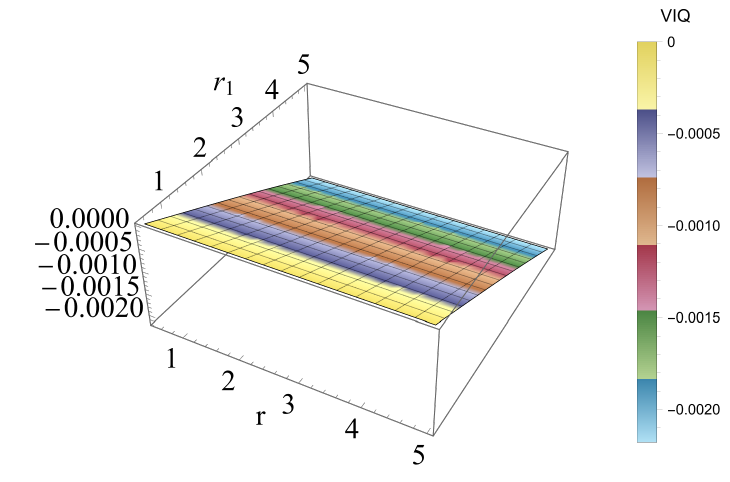}
    \caption{VIQ for MIT bag model shape function-I (first), shape function-II (second), and GCCG constant redshift (third) for the same value of constants.}
    \label{fig:a2}
\end{figure}
In Fig. \ref{fig:a2}, the characteristics of the volume integral $Iv$ for the MIT bag model and GCCG WHs are illustrated. It is noted that as $r_1$ approaches $r_0$, $Iv$ tends to zero. The results depicted in these figures confirm that our solutions adhere to this expected behavior. Consequently, this demonstrates the possibility of spacetime structures that include traversable wormholes, which can be maintained with minimal amounts of exotic matter. Moreover, by selecting an appropriate geometry for the wormhole, the total quantity of matter that violates the ANEC can be minimized.

\section{Discussions and Conclusions}
\label{sec9}
In this paper, we explored WH solutions in $f(Q)$ gravity under different scenarios. Firstly, we examined the impact of the MIT bag model, and then we looked at the effects of generalized cosmic chaplain gas on WH geometry. In the first scenario, we initially considered the isotropic EoS relation along with the logarithmic redshift function to calculate the shape function. However, we found that the obtained shape function did not satisfy the asymptotic flatness condition. Additionally, we analyzed the Bag parameter and found that at the throat of the WH, the value of $B(r)$ was $140\, \text{MeV/fm}^3$, which falls within the range of $[41.58,\, 319.13]\, \text{MeV/fm}^3$ \cite{mit1,mit2}. It is worth noting that in the case of GR, the value is on the order of $10^6$, which is significantly higher than the relevant range of $[41.58,\, 319.13]\, \text{MeV/fm}^3$, making it inconsequential. We also observed that as $\omega\to 0$, the bag pressure $B(r)$ increases and approaches its upper limit. This suggests that during the thermodynamic phase transition from the matter-dominated to the radiation-dominated era, our model indicates enhanced quark interactions within the WH.\\
Further, we extend our analysis for anisotropic MIT bag model EoS and obtained bag parameter $B(r)$. Subsequently, we investigated the behavior of the bag parameter using different shape functions. It is observed that for shape function (I), the value of the bag parameter $B(r)$ at the throat is $160\, \text{MeV/fm}^3$, which falls within the interval \cite{mit1,mit2}. On the other hand, for shape function (II), the value slightly exceeds the required interval. However, for both shape functions, the value of $B(r)$ does not meet the necessary interval within the framework of GR. It is also noted that for each case, as we increase $\omega$, the bag parameter $B(r)$ exhibits a consistently decreasing behavior near the throat. Furthermore, far from the throat, it monotonically decreases, indicating minimal interaction between the quark matter. Additionally, we analyze the energy density, NEC, and SEC near the WH throat for each shape function. We find that NEC is violated in each case. Furthermore, we observe the impact of free parameters on the violation of NEC. As already discussed in the introduction, wormhole solutions have been studied in $f(R, T)$ gravity \cite{Tayde1}, focusing on SQM characterized by both isotropic and anisotropic pressures, and this work has been further extended in $f(Q,T)$ gravity with constant redshift function \cite{Tayde2}. In this study, we explored wormhole geometry within $f(Q)$ gravity, employing the MIT bag model alongside various EoS formulations with non-constant redshift functions.\\
In the second phase of this paper, we investigated wormhole solutions supported by the generalized Chaplygin gas in $f(Q)$ gravity. The analysis was conducted under three different forms of redshift functions. We derived three distinct forms of shape functions by employing the generalized Chaplygin gas under these redshift functions. We observed that the shape functions satisfied the flare-out condition with the appropriate choice of parameters. It was also noted that the negative behavior of $\delta$ contributed to meeting the shape function criteria for each case. Energy conditions were checked for each scenario, and it was found that $\rho+p_r$ is violated, indicating a violation of the averaged null energy condition. Moreover, we observed that increasing the negative values of $\delta$ resulted in a more significant contribution to the violation of NEC. This violation is required for the existence of any wormhole solution. Interestingly, previous studies by Lobo \cite{lobo} on Chaplygin traversable wormholes suggested that the Chaplygin gas needs to be confined around the wormhole throat neighborhood. This notion was further explored in \cite{chakraborty}, where modified Chaplygin gas led to the deduction that modified Chaplygin wormholes may occur naturally and be traversable. Further, in the context of $f(R,T)$ gravity, wormhole solutions have been explored with two different forms of Chaplygin gas \cite{Emilio}. Our results align with these findings, as our solutions satisfy wormhole stability and traversability criteria.\\
We further discussed VIQ analytically. This effectively tells us the amount of exotic matter required to violate the ANEC. Our analysis found that a small amount of exotic matter is necessary for a traversable wormhole. Since wormhole geometries may be unstable for non-traversable wormholes, we investigated the stability of our wormhole solution using the generalized TOV equation. The result showed that all the physical forces acting on the wormhole effectively cancel out, providing stable wormhole solutions.\\
Thus, it is safe to conclude that our obtained results are physically viable in the context of $f(Q)$ gravity. Also, the procedure demonstrated here can be used with other theories of gravity to find new wormhole solutions containing strange matter and to restrict them based on energy conditions. It would be valuable to calculate wormhole solutions that follow the EoS of the MIT Bag Model in non-commutative geometry using the method presented in \cite{Sharif}. We aim to share more about these concepts in the near future.

\section*{Data availability} No new data were generated or analyzed in support of this research.

\section*{Acknowledgement}
SP \& PKS  express gratitude to the National Board for Higher Mathematics (NBHM) under the Department of Atomic Energy (DAE), Government. of India, for providing financial assistance to conduct the Research project No.: 02011/3/2022 NBHM(R.P.)/R \& D II/2152 Dt.14.02.2022. ZH acknowledges the Department of Science and Technology (DST), Government of India, New Delhi, for awarding a Senior Research Fellowship (File No. DST/INSPIRE Fellowship/2019/IF190911).

\section*{\textbf{Appendix}}

\begin{eqnarray}
\mathcal{G}_1=\Gamma \left(\frac{1}{(1-\eta )^{-\delta }+\eta },-\frac{2 \chi (1-\eta )^{\delta }}{\eta  r_0 (1-\eta )^{\delta }+r_0}\right)\nonumber\\
    \mathcal{G}_2=\Gamma \left(\frac{1}{(1-\eta )^{-\delta }+\eta }-3,-\frac{2 \chi (1-\eta )^{\delta }}{\eta  r_0 (1-\eta )^{\delta }+r_0}\right)\nonumber\\
    \mathcal{I}_1=\text{ExpIntegral}\Big[{1-\frac{1}{(1-\eta )^{-\delta }+\eta }};-\frac{2 \chi (1-\eta )^{\delta }}{\eta  r (1-\eta )^{\delta }+r}\Big]\nonumber\\
    \mathcal{I}_2=\text{ExpIntegral}\Big[{4-\frac{1}{(1-\eta )^{-\delta }+\eta }}-\frac{2 \chi (1-\eta )^{\delta }}{\eta  r (1-\eta )^{\delta }+r}]\nonumber   
\end{eqnarray}

\begin{eqnarray}
   \nonumber\\&&\hspace{0cm} d_1=\frac{1}{d_2}\Big(e^{\frac{(1-\eta )^{\delta }  \log (\frac{(r+r_0)^2}{r})}{\Gamma_2}}r^{\frac{(1-\eta )^{\delta }}{\eta  (1-\eta )^{\delta }+1}+1}(r_0)^{-\frac{2 (1-\eta )^{\delta }}{\eta  (1-\eta )^{\delta }+1}}\nonumber\\&&\hspace{1cm}\big(\beta  r^2 \mathcal{F}_3\big[(\eta ^2 (1-\eta )^{2 \delta }+2 \eta  (1-\eta )^{\delta }-(1-\eta )^{2 \delta }+1\big]\nonumber\\&&\hspace{1cm}+4 \alpha \mathcal{F}_4 (1-\eta )^{\delta } \left(3 \eta  (1-\eta )^{\delta }+(1-\eta )^{\delta }+3\right)\big)\Big) \nonumber
\end{eqnarray}
Where $$d_2=2 \alpha  \left(\eta  (1-\eta )^{\delta }+(1-\eta )^{\delta }+1\right) \left(3 \eta  (1-\eta )^{\delta }+(1-\eta )^{\delta }+3\right)$$

\begin{eqnarray}
&&\hspace{0cm}\mathcal{F}_2=\, _2F_1\Big[\frac{2 (1-\eta )^{\delta }}{\eta  (1-\eta )^{\delta }+1},\frac{3 \eta  (1-\eta )^{\delta }+(1-\eta )^{\delta }+3}{\eta  (1-\eta )^{\delta }+1};\nonumber\\&&\hspace{1cm}\frac{4 \eta  (1-\eta )^{\delta }+(1-\eta )^{\delta }+4}{\eta  (1-\eta )^{\delta }+1};-1\Big]\nonumber\\ 
&&\hspace{0cm} \mathcal{F}_2 = \, _2F_1\Big[\frac{\eta  (1-\eta )^{\delta }+(1-\eta )^{\delta }+1}{\eta  (1-\eta )^{\delta }+1},\frac{2 (1-\eta )^{\delta }}{\eta  (1-\eta )^{\delta }+1}+1;\nonumber\\&&\hspace{1cm}\frac{2 \eta  (1-\eta )^{\delta }+(1-\eta )^{\delta }+2}{\eta  (1-\eta )^{\delta }+1};-1\Big]\nonumber\\
&&\hspace{0cm}\mathcal{F}_3=\, _2F_1\big[\frac{2 (1-\eta )^{\delta }}{\eta  (1-\eta )^{\delta }+1},\frac{3 \eta  (1-\eta )^{\delta }+(1-\eta )^{\delta }+3}{\eta  (1-\eta )^{\delta }+1};\nonumber\\&&\hspace{1cm}\frac{4 \eta  (1-\eta )^{\delta }+(1-\eta )^{\delta }+4}{\eta  (1-\eta )^{\delta }+1};-\frac{r}{r_0}\big]\nonumber\\
&&\hspace{0cm}\mathcal{F}_4=\, _2F_1\Big[\frac{\eta  (1-\eta )^{\delta }+(1-\eta )^{\delta }+1}{\eta  (1-\eta )^{\delta }+1},\frac{2 (1-\eta )^{\delta }}{\eta  (1-\eta )^{\delta }+1}+1;\nonumber\\&&\hspace{1cm}\frac{2 \eta  (1-\eta )^{\delta }+(1-\eta )^{\delta }+2}{\eta  (1-\eta )^{\delta }+1};-\frac{r}{r_0}\Big]\nonumber
\end{eqnarray}

\end{document}